\documentclass{aa}  

\usepackage{graphicx}
\usepackage{txfonts}
\usepackage{hyperref}
\usepackage{xcolor}
\usepackage{upgreek}
\usepackage{wasysym}

\usepackage{booktabs}
\begin{document}

    \title{MeerKAT observations of the spiral galaxy NGC\,2997 in the S band.}
    \subtitle{Detection of high dynamo modes}

   \titlerunning{MeerKAT S-band observations of NGC\,2997}
   \authorrunning{A. Damas-Segovia et al.}
   
   \author{A. Damas-Segovia
          \inst{1},
          R. Beck
          \inst{1},
          S. A. Mao
          \inst{1},
          A. Basu
          \inst{1,2},
          S. Sridhar
          \inst{1,3},
          E. Barr
          \inst{1},
          A. Brunthaler
          \inst{1},
          S. Buchner
          \inst{4},
          F. Camilo
          \inst{4},
          W. Cotton
          \inst{4,5},
          A. M. Jacob
          \inst{1,6},
          C. Kasemann
          \inst{1},
          H.-R. Kl\"ockner
          \inst{1},
          M. Kramer
          \inst{1,7},
          I. Rammala-Zitha
          \inst{1},
          S. Ranchod
          \inst{1},
          M. R. Rugel
          \inst{1,8,9},
          O. Smirnov
          \inst{4,10,11},
          J. D. Wagenveld
          \inst{1},
          G. Wieching
          \inst{1},
          O. Wucknitz
          \inst{1}.}

   \institute{Max-Planck-Institut f\"ur Radioastronomie, Auf dem H\"ugel 69, 53121 Bonn, Germany\\
              \email{adamas@mpifr-bonn.mpg.de}
         \and
            Th\"uringer Landessternwarte, Sternwarte 5, 07778 Tautenburg, Germany
         \and
            SKA Observatory, Jodrell Bank, Lower Withington, Macclesfield, SK11 9FT, United Kingdom 
         \and 
          South African Radio Astronomy Observatory, Liesbeek House, River Park, Cape Town 7705, South Africa
         \and
         National Radio Astronomy Observatory, 520 Edgemont Road, Charlottesville, VA 22903, USA 
         \and
         William H. Miller III Department of Physics \& Astronomy, Johns Hopkins University, 3400 North Charles Street, Baltimore, MD 21218, USA
         \and
         Jodrell Bank Centre for Astrophysics, Department of Physics and Astronomy, The University of Manchester, Manchester M13 9PL, UK
         \and
         National Radio Astronomy Observatory, P.O. Box O, 1003 Lopezville Road, Socorro, NM 87801, USA
         \and
         Center for Astrophysics | Harvard \& Smithsonian, 60 Garden Street, Cambridge, MA 02138, USA 
         \and
         Centre for Radio Astronomy Techniques and Technologies (RATT), Department of Physics and Electronics, Rhodes University, Makhanda, 6140, South Africa
         \and
         Institute for Radioastronomy, National Institute of Astrophysics (INAF IRA), Via Gobetti 101, 40129 Bologna, Italy
         }

   \date{Received ***; accepted ***}

% \abstract{}{}{}{}{}
% 5 {} token are mandatory
 
  \abstract
  %context heading (optional)
   {}
  %aims heading (mandatory
   { We seek to exploit the expanded observational range of the MeerKAT radio telescope with the new S-band receivers (2.0 -- 2.8 GHz). To showcase its enhanced capabilities, we conducted new S-band observations of the galaxy NGC\,2997 in full polarization. The S band is ideal for studying magnetic fields in spiral galaxies due to the weak Faraday depolarization.}
  %methods
   {For the data calibration procedure, we utilized the Max Planck MeerKAT Galactic Plane Survey (MMGPS) pipeline, capable of performing full-Stokes calibration, self-calibration, and imaging of MeerKAT data. Performing a rotation measure (RM) synthesis allowed us to measure Faraday RMs in the galaxy, a signature of regular magnetic fields. A fast Fourier transform (FFT) algorithm was used to study the various azimuthal modes found in the RM data of the galaxy.}
  %results
   {The final radio maps in total and polarized intensity are characterized by root mean square (rms) noise of $11\,\mu\rm{Jy}\,\rm{beam}^{-1}$ at a resolution of about $4\arcsec$, enabling a detailed examination of the galaxy. The total radio intensity map reveals the spiral arm structure associated with star-forming regions and the inner ring around the nucleus. Additionally, the galaxy exhibits strong polarized emission indicative of a large-scale ordered magnetic field. This ordered field traces the optical spiral structure of the disk in the south, whereas we see larger pitch angles in the north. The RM synthesis analysis indicates the direction of the magnetic field along the line of sight throughout the entire disk. Leveraging the sensitivity and high resolution provided by  MeerKAT's S-band capability, this study achieves an unprecedented level of detail of the magnetic field structure. Our sector-based analysis of the RMs across azimuthal regions reveals the existence of modes of the large-scale magnetic field in NGC\,2997. The variations in the RM values along the azimuthal angle reveal smoothly changing phase shifts between the rings, without the previously reported field reversal at about 3\,kpc radius between the central region and disk. Further refinement approaches would involve computing the RMs, while correcting for the inclination of the disk and considering the position angle of the major axis. In this work, for the first time, a Fourier analysis has been applied to RM data averaged in sectors of rings in the disk plane of a spiral galaxy.}
  %conclusions
   {Our Fourier analysis of the RM map shows three different large-scale field modes detected in the disk of NGC\,2997. After applying a geometric modification, even multiples of the first mode were detected, as predicted from theoretical studies of dynamo action in a spiral galaxy with symmetric spiral structure. Our new method opens up new possibilities for investigating magnetic fields in spiral galaxies.}


   \maketitle

\section{Introduction}

Magnetic fields are present at every scale in the Universe. Cosmic magnetism is explored across stellar physics, galaxy evolution, and cosmological observations \citep{Beck_2015}. Since the early days of radio astronomy, linear polarization observations have played a pivotal role in advancing our understanding of cosmic magnetism.

In the exploration of galactic magnetic fields, the inclination of the galaxy disk with respect to the line of sight emerges as a key factor influencing the observable field structure in full-Stokes radio observations. This phenomenon is particularly evident in two extreme scenarios:\ edge-on and face-on spirals, each revealing distinctive components of the ordered magnetic field.

In the case of edge-on spiral galaxies, characterized by an inclination at or close to $90\degr$, linearly polarized radio synchrotron emission reveals the strength of the ordered magnetic field and its orientation (ambiguous by $\pm$ $180\degr$), marked by two discernible structures \citep[see CHANG-ES project,][]{Irwin_2012}. The first component is a plane-parallel magnetic field intricately associated with the galaxy's disk and linked to the interstellar medium (ISM) within the galaxy. The second component encompasses a large-scale ordered magnetic field that extends into the galactic halo, spanning kiloparsec (kpc) scales toward the outer regions of the galaxy. Notably, the halo field often adopts an X-shaped structure and is conventionally correlated with cosmic ray electrons (CREs) that escape from the galactic disk due to a heightened star formation rate (SFR). Faraday rotation measures (RMs) serve as a valuable tool for probing the strength and direction of the large-scale field (i.e., the ``regular field'') along the line of sight. Positive and negative RM values correspond to magnetic fields pointing toward and away from the observer, respectively. In the context of edge-on galaxies, RMs can reveal a highly complex magnetic field, particularly in regions distant from the galactic disk \citep{Mora_2019,Krause_2020}. Edge-on galaxies are the perfect laboratory to study the interconnection between disk and halo through the analysis of CRE transport models \citep{Heesen_2016}.

Conversely, galaxies at a nearly face-on view exhibit a distinctive spiral-arm pattern in their polarized radio intensity, primarily emanating from the inter-arm regions \citep{Beck_2015_IC}, referred to as magnetic spiral arms. The origin of regular spiral fields is thought to be the large-scale dynamo \citep{Beck_1996}, describing the magnetic field as a combination of sinusoidal functions with azimuthal angle in the galaxy plane (``mode'').
Enhanced star formation in the ISM of spiral arms generates fast outflows that can suppress the action of the large-scale dynamo \citep{Chamandy_2015}, leaving regular fields mostly to inter-arm regions. Furthermore, turbulence within the ISM of spiral arms leads to the tangling of ordered fields, which reduces the observed polarized intensity.
Analyzing the distribution of RMs within the disk of a face-on galaxy provides insights into the various azimuthal modes of the large-scale magnetic field \citep{Beck_2019}.

The majority of observed galaxies exhibit some degree of inclination, impacting the apparent size of their disks in the plane of the sky. Consequently, any analysis should consider projection effects when measuring any physical characteristic of the galaxy. 

In magnetic field studies, inclination plays a fundamental role in how observers perceive the magnetic field along the line of sight. The RM values are directly proportional to the average magnetic field component parallel to the line of sight ($B_ \parallel$) and, thus, the disk's inclination affects the RM pattern. In face-on galaxies,  purely due to projection effects, only plane-perpendicular field components lead to RMs. In mildly inclined galaxies, both plane-parallel and vertical fields contribute to RMs.

NGC\,2997 (see details in Table \ref{galaxy}) has a comparable star formation rate, akin to that of the Milky Way, and it lacks discernible signs of strong interaction. Despite the recent revelation of a substantial population of dwarf galaxies in its proximity \citep{Fan_2023}, there are no large companions in its vicinity. The disk is inclined by about $40^\circ$ and exhibits a symmetrical shape in optical observations, revealing a lagging halo in neutral hydrogen observations \citep{Hess_2009}. This lagging halo is likely a consequence of a galactic fountain scenario, wherein supernova events expel material from the ISM into the halo. These pronounced effects may be indicative of an influence on the magnetic field within the disk.

The total and polarized radio emission of NGC\,2997 was previously observed by \cite{Han_1999} with the Very Large Array (VLA) and the Australia Telescope Compact Array (ATCA). At 4.86\,GHz and 8.46\,GHz, the total intensity emerges mainly from the regions of the optical spiral arms. The polarized intensity at 4.86\,GHz at $15\arcsec$ resolution is generally stronger along the inner edge of the optical arms, but also from two inter-arm regions that may form the brightest part of two "magnetic arms." The map of RM between 2.3\,GHz and 4.86\,GHz at $25\arcsec$ resolution indicates that the regular field in the northern spiral arm points inwards toward the galaxy's center, while the large-scale pattern of the regular field could not be determined due to the lack of sensitivity. In the central region, between a radius of 0.5\,kpc and 2.0\,kpc in the galaxy's plane, an RM between 4.86\,GHz and 8.46\,GHz at $11\arcsec$ resolution varies sinusoidally along the azimuthal angle, indicative of an axisymmetric spiral pattern of the regular field. Notably, the field direction points outwards, opposite to the case seen for the northern spiral arm.

Polarization observations in the S band (2--3\,GHz) are crucial for studying magnetic fields in spiral galaxies. In this frequency band, Faraday depolarization effects are less severe than at lower frequencies (e.g., the L band) and thanks to the steep synchrotron spectrum, we can detect larger flux densities compared to higher frequencies such as the C band or X band. The first complete polarization map of a spiral galaxy (M\,31) was obtained in the S band \citep{Beck_1980}.
Previous high-resolution observations using the S band demonstrated that this frequency range is fundamental to investigate both plane-parallel and vertical magnetic fields in spiral galaxies (e.g., NGC\,628 \citep{Mulcahy_2017} and M\,51 \citep{Kierdorf_2020}). Therefore, the primary motivation of the current study is to use the new capabilities of the MeerKAT  telescope \citep{Jonas_2016} to enhance our understanding of the magnetic properties of NGC\,2997. This research will serve as a test case for future studies involving magnetic fields in other galaxies.

\begin{table}
        \centering
        \caption{Properties of NGC\,2997.}
        \begin{tabular}{lccr}
            \toprule
                Right ascension & 09h 45m 38.8s  & 1\\
                Declination & $-$31\degr 11\arcmin 28\arcsec & 1\\
            Hubble type & SAB(rs)c & 2\\
            Distance & 14.8\,Mpc & 3\\
            Major axis (D$_{25}$)& 8.9\arcmin  & 1\\
            Minor axis & 6.8\arcmin & 1\\
            Inclination & 40\degr & 4 \\
            Position angle & 110\degr & 4\\
            Star formation rate & 4.4 $\rm{M}_\odot$\,yr$^{-1}$ & 5\\
                \bottomrule
        \end{tabular}
\tablebib{(1) NASA/IPAC Extragalactic Database; (2) \citet{Healey_2007}; (3) \citet{Pisano_2011}, assuming $H_0 = 72\,\rm{km}\,\rm{s}^{-1}\,\rm{Mpc}^{-1}$; (4) \citet{Milliard_1981} (kinematic); (5) \citet{Leroy_2021}.

}
\label{galaxy}
\end{table}

In the subsequent sections of this paper, we show the characteristics of the observations and present our data reduction strategy implemented via our pipeline (Section \ref{sec:observations}). The results are detailed in Section \ref{sec:results}, where we provide maps of the total radio intensity, polarized intensity, polarization orientations, spectral index, and Faraday RMs. A comprehensive discussion follows in Section \ref{sec:discussion}, providing a dedicated analysis of the RM distribution of this galaxy. Finally, we present our conclusions in Section \ref{sec:conclusions}.

\section{Observations and data reduction}
\label{sec:observations}

\begin{table}
        \centering
        \caption{Parameters of the radio continuum observations.}
        \begin{tabular}{lccr}
                \toprule
                Number of antennas & 57\\
                Central frequency & 2406.25 MHz\\
            Bandwidth & 875.00 MHz\\
            Channels & 4096 \\
            Channel width & 213.623 kHz\\
            Bandpass calibrator & PKS 0408-658 \\
            Gain calibrator & J0825-5010 \\
            Polarization calibrator & 3C138 \\
            Leakage calibrator & J0825-5010 \\
            Date & 2nd and 5th April 2023\\ 
            Total observing time per day & 3.5 hr \\
            On-source observing time & 2.8 hr\\
            
                \bottomrule
        \end{tabular}
    \label{observations}
\end{table}

\begin{figure*}
    \centering
    \begin{minipage}{0.5\textwidth}
        \centering
        \includegraphics[width=\textwidth]{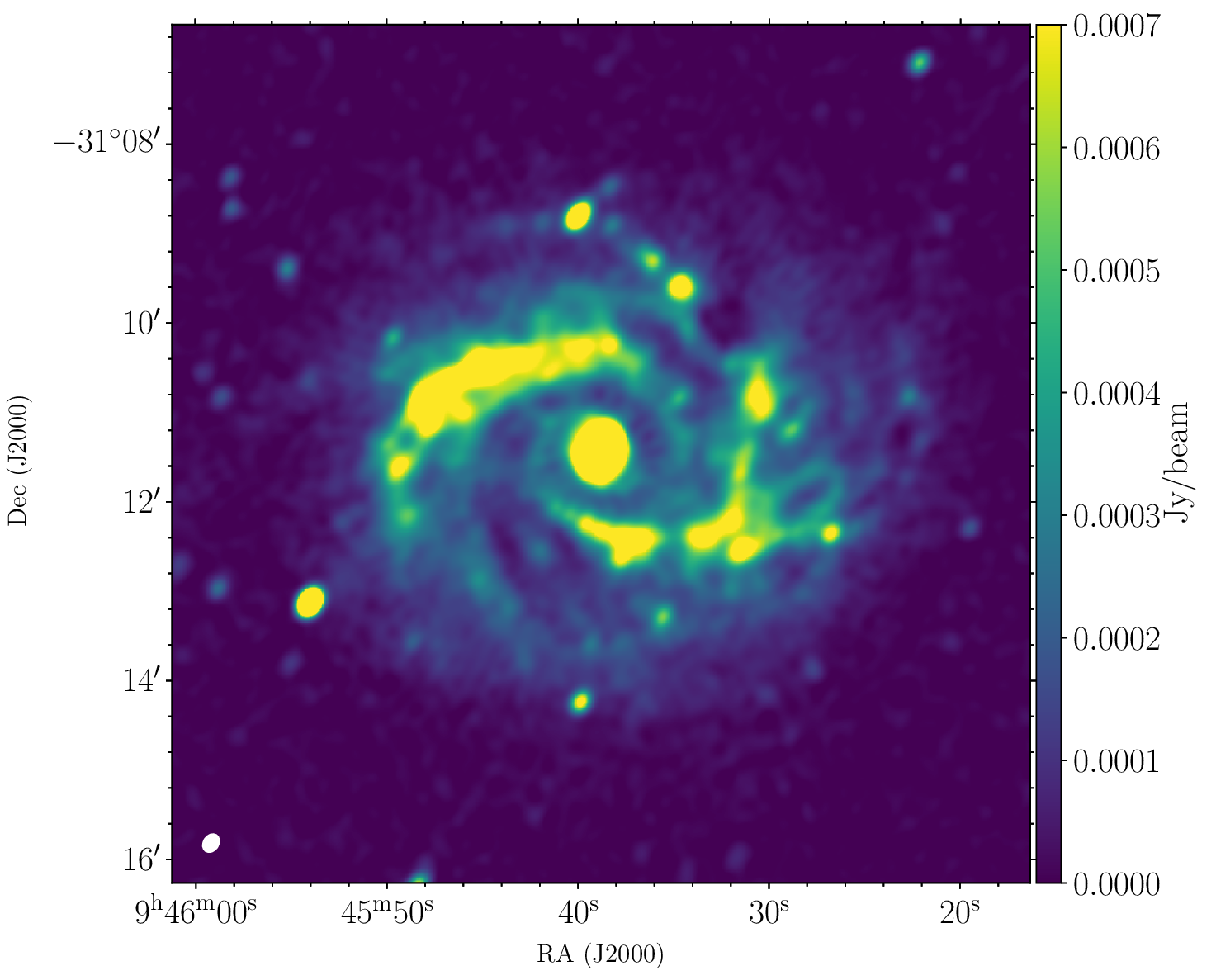}
        \label{fig:total_I}
    \end{minipage}\hfill
    \begin{minipage}{0.5\textwidth}
        \centering
        \includegraphics[width=\textwidth]{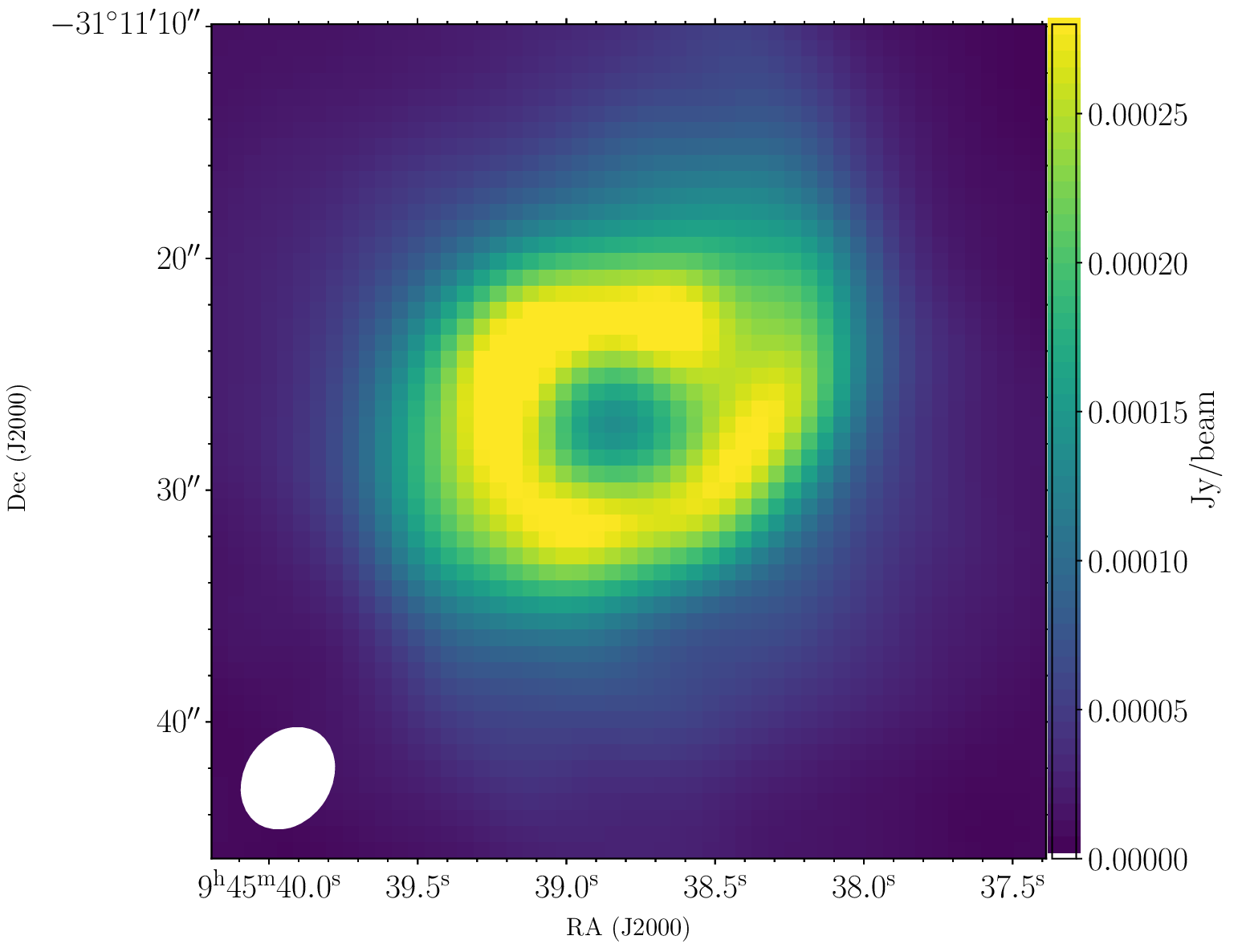}
    \end{minipage}
    \caption{Total radio intensity maps of NGC\,2997 detected in the S band with MeerKAT. The scales are in Jy\,beam$^{-1}$. {\bf Left:} Total radio intensity applying a robust 0 weighting during the cleaning process. This map shows an rms of $11\,\mu\rm{Jy\,beam}^{-1}$ at a resolution of 12.8" x 9.8", with a position angle of -33.5\degr. The beam size is shown at the bottom left corner. {\bf Right:} Total radio intensity applying robust -1 weighting during the cleaning process to show the inner ring of the central region of the galaxy. The rms of this map is $1.5\,\mu\rm{Jy\,beam}^{-1}$ at a resolution of 4.6" x 3.7", with a position angle of -34\degr.}
    \label{fig:total_I}
\end{figure*}

The observations of NGC\,2997 (project code: SSV-20230209-AD-01) were performed with the MeerKAT telescope array run by SARAO using the S1 sub-band of the new S-band receiver system covering the frequency range 1.968 -- 2.843\,GHz in full polarization mode. More details of this system are presented in \citet{Ranchod_2025}.
The total duration of 7\,hours (hr) for these observations, including calibration overheads, were split over two days of 3.5\,hr each. The total on-source time on NGC\,2997 was 2.8\,hr, using 57 and 56 operational antennas during these days. The total time on-source was estimated to improve the sensitivity of previous observations of this object by a factor of about 2 \citep{Han_1999}. The 875\,MHz bandwidth was split into 4096 frequency channels of 213.623\,kHz. In Table~\ref{observations}, we summarize the observational details and the different calibrators used.

These data were calibrated using a fully automated pipeline developed to analyze data from the MPIfR MeerKAT Galactic Plane Survey \citep[MMGPS;][]{Padmanabh}. The MMGPS imaging pipeline was developed in-house as a collaborative effort between the Max-Planck-Institut f\"ur Radioastronomie (MPIfR), Bonn, and the Th\"uringer Landessternwarte (TLS), Tautenburg. The pipeline is a modular collection of several analysis packages, containerized in \texttt{Singularity}, which performs specific tasks. A full Stokes calibration including delay, bandpass, complex gain, leakage, polarization angle, and basic flagging was performed using the Common Astronomy Software Application \cite[\texttt{CASA version 6.4};][]{mcmullin_2007}, while rigorous flagging was performed using the \texttt{Tricolour} package\footnote{\url{https://github.com/ratt-ru/tricolour}}. Full-Stokes imaging was performed using the \texttt{WSClean} package \citep[version\,3.1;][]{offringa-wsclean-2014,offringa-wsclean-2017}, while self-calibration was performed with Stokes\,$I$ images and solutions using CASA.

The various parameter inputs in each of the packages were controlled by a user provided configuration file for the different rounds of self-calibration.
The flux density scale in these observations were transferred to NGC\,2997 based on the model of PKS\,0408-658, which has a flux density of $7.829\pm0.002\,\rm{Jy \, beam^{-1}}$ at 2.4\,GHz. This model includes the spectral index of this calibratior.

The on-axis leakage level in linear polarization was estimated from the ratio between the flux densities of the leakage calibrator J0825-5010 by integrating over an area of 8\arcsec\ of Stokes Q and U and total intensity, I. The polarization fractions were found to be $\rm{Q/I}=0.0009$ and $\rm{U/I}=0.0005$, so that the on-axis polarization leakage level is less than 0.11\%.

After flagging and calibration, four adjacent channels were averaged, each having 854.5\,kHz in width. For the self-calibration procedure, these channel-averaged data were split into 16\,sub-bands (spectral windows) of about 47.85\,MHz in width containing 56 channels each, namely, $\Delta\nu/\nu = 0.036\textrm{--}0.047$.
Each sub-band was independently self-calibrated and since $\Delta\nu/\nu \le 0.05$ across all the sub-bands, we did not include any higher order frequency dependence of the flux density variation within each sub-band during imaging. For these observations, we performed three rounds of "phase-only" self-calibration and one round of "amplitude and phase" self-calibration. The self-calibration timescales used were $30, 10, 5$, and $300$ seconds.
Imaging in Stokes I, Q, and U was performed using different weighting values of the robust parameter in \texttt{wsclean} to study different scales of the radio emission in this galaxy. 
The primary beam correction was performed using image-based models from the KATBeam\footnote{\url{https://github.com/ska-sa/katbeam}} software package, outside the pipeline workflow.

\iffalse

\fi
The polarization products (intensity, angle, and RM) were obtained using RM synthesis \citep{Brentjens_2005}.
Image cubes of 896 channels were produced for Stokes $Q$ and $U$, with each channel having a width of 854\,kHz. The frequency range for this analysis was $2.014\,\rm{GHz} - 2.788\,\rm{GHz}$.
The resulting Faraday spectrum (i.e., polarized intensity as a function of Faraday depth) at each 5-sigma pixel of the map was estimated to have a resolution of 365 rad/m², measured from the full width half maximum (FWHM) of the RM transfer function.
A map of Faraday RM was computed from the highest peak in each Faraday spectrum. No components wider than our Faraday resolution were found.
Errors of the RM synthesis products were computed using the equations published in Appendix A of \cite{Brentjens_2005}.

Dividing Stokes $Q$ and $U$ by the total power emission could minimize the effects of the spectral index on polarized flux density and RM. We did not use this technique in our study due to the expected contribution of the thermal component, which would introduce significant uncertainty in the final non-thermal flux density of the data.

\section{Results}
\label{sec:results}

In this section, we present the main results of the calibration and imaging process.

\subsection{Total intensity}

\begin{table}
        \centering
        \caption{Total flux density measurements of NGC\,2997.}
        \begin{tabular}{lccr}
            \toprule
            {Frequency (GHz)}& {Flux density (Jy)} & {Reference}\\
            \midrule
                8.460 & ${0.034\pm0.004}^*$ & \cite{Han_1999}\\
                5.010 & $0.092\pm0.010$ & \cite{Whiteoak_1970}\\
            4.860 & ${0.067\pm0.011}^*$ & \cite{Han_1999}\\
            4.850 & $0.141\pm0.014$& \cite{Wright_1996}\\
            2.373 & ${0.160\pm0.010}^*$ & \cite{Han_1999}\\
            2.741 & $0.190\pm0.001$ & spw 0 of this study\\
            2.644 & $0.198\pm0.001$ & spw 1 of this study\\
            2.547 & $0.207\pm0.001$ & spw 2 of this study\\
            2.450 & $0.217\pm0.001$ & spw 3 of this study\\
            2.406 & $0.227\pm0.007$ & this study (average)\\
            2.353 & $0.227\pm0.001$ & spw 4 of this study\\
            2.256 & $0.238\pm0.001$ & spw 5 of this study\\
            2.159 & $0.249\pm0.001$ & spw 6 of this study\\
            2.062 & $0.262\pm0.001$ & spw 7 of this study\\
            1.543 & $0.255\pm0.020$ & \cite{Han_1999}\\
            1.490 & $0.290$ & \cite{Condon_1987}\\
            1.400 & ${0.235\pm0.010}^*$& \cite{Condon_1998}\\
            1.272 & $0.367\pm0.008$& \cite{Kodilkar_2011}\\
            0.616 & $0.730\pm0.022$& \cite{Kodilkar_2011}\\
            0.332 & $1.134\pm0.111$& \cite{Kodilkar_2011}\\
                \bottomrule
\end{tabular}
\label{flux_literature_table}
\tablefoot{*: Measurements with insufficient detection of extended emission.
}
\end{table}

\begin{figure}
\centering

        \includegraphics[width=1.0\columnwidth]{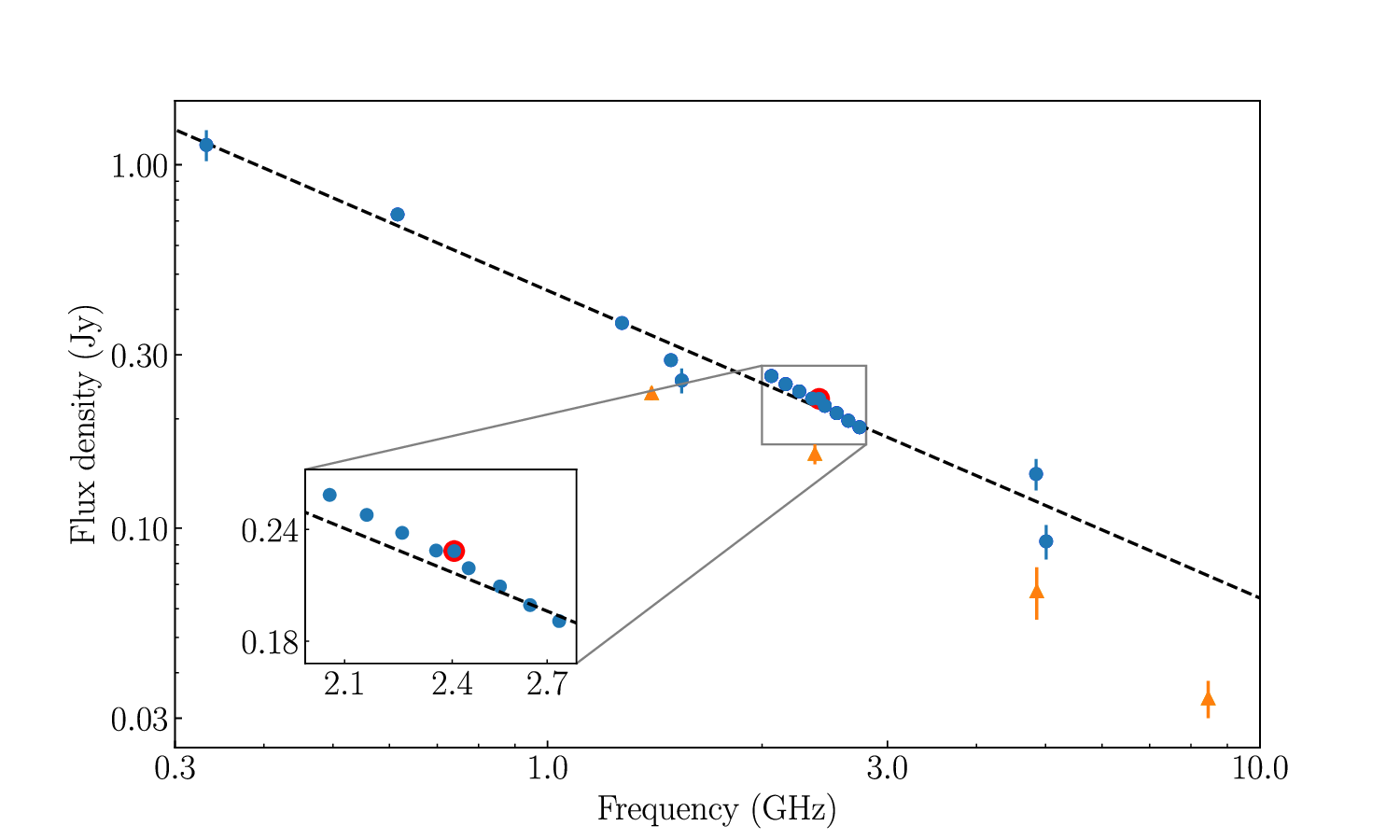}
        \caption{Total flux density of different observations of NGC\,2997 compared to the MeerKAT S-band observations. The points represent observations listed in Table \ref{flux_literature_table}. Blue circles in this plot correspond to studies that fully recover the extended emission of the source. These data points were used to estimate the spectral index of the total flux densities. This linear fit is represented by the black dashed line. Orange triangles represent observations that did not fully recover the extended emission of the source. These data points were not included in the fit. \textbf{Inset:} Total intensity emission (robust 0 weighting) in 8 spectral windows. The maps in each spectral window were smoothed to the resolution at the lowest frequency and corrected with the corresponding primary beam model. The blue circles represent the flux density integrated over the whole galaxy, integrated out to a radius of 250\arcsec. The red circle represents the integrated flux density over the whole bandwidth. The rms per spectral window is about $40\,\mu$Jy\,beam$^{-1}$
        which gives a flux density error of about 1\,mJy, similar to the symbol size. The general uncertainty of the absolute calibration scale of about 3\% is not included in this plot. }
    \label{fig:flux_literature}
\end{figure}

\begin{figure*}
    \centering
    \begin{minipage}{0.5\textwidth}
        \centering
        \includegraphics[width=\textwidth]{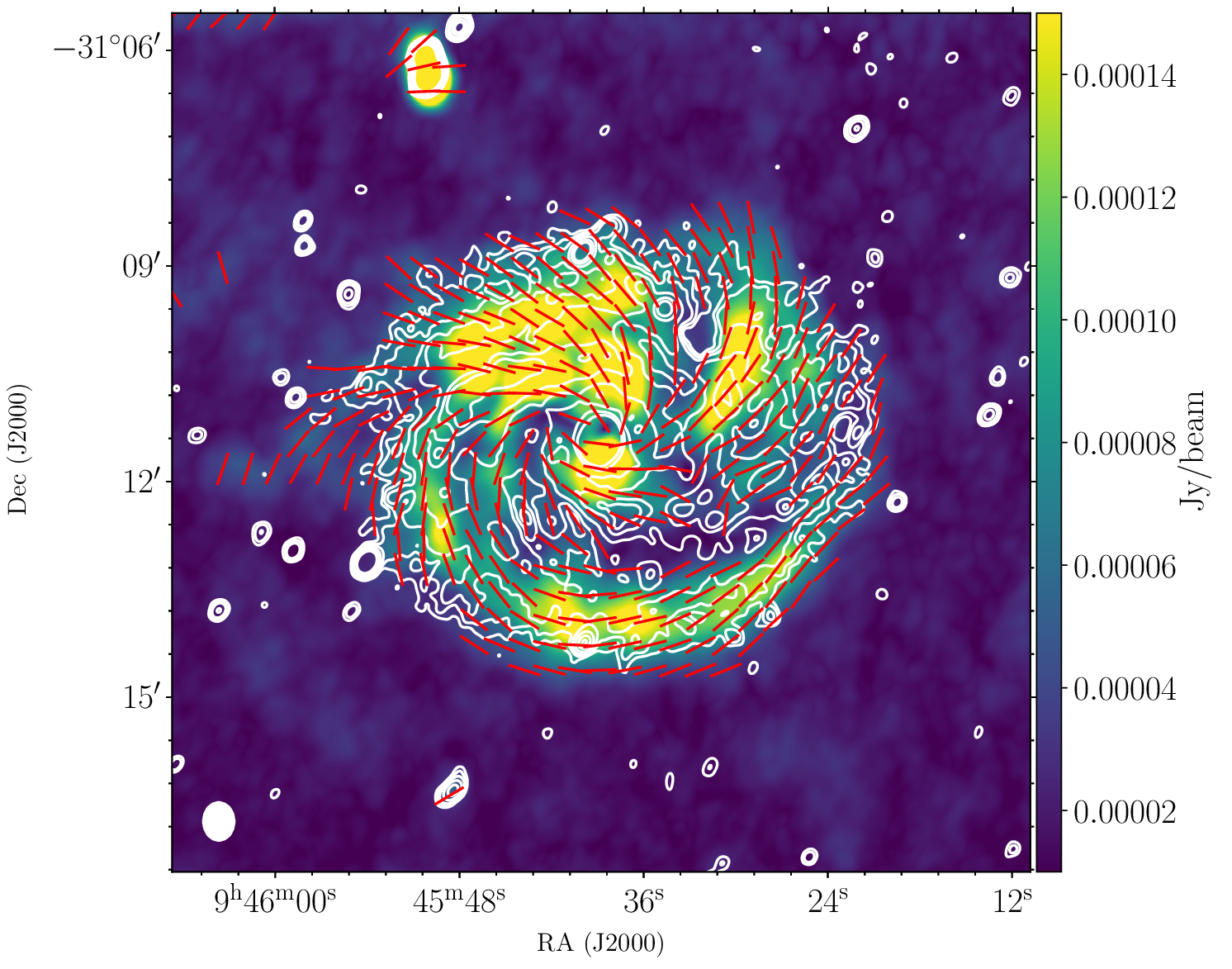}
        %\caption{Caption for the first figure}
        \label{fig:PI}
    \end{minipage}\hfill
    \begin{minipage}{0.5\textwidth}
        \centering
        \includegraphics[width=\textwidth]{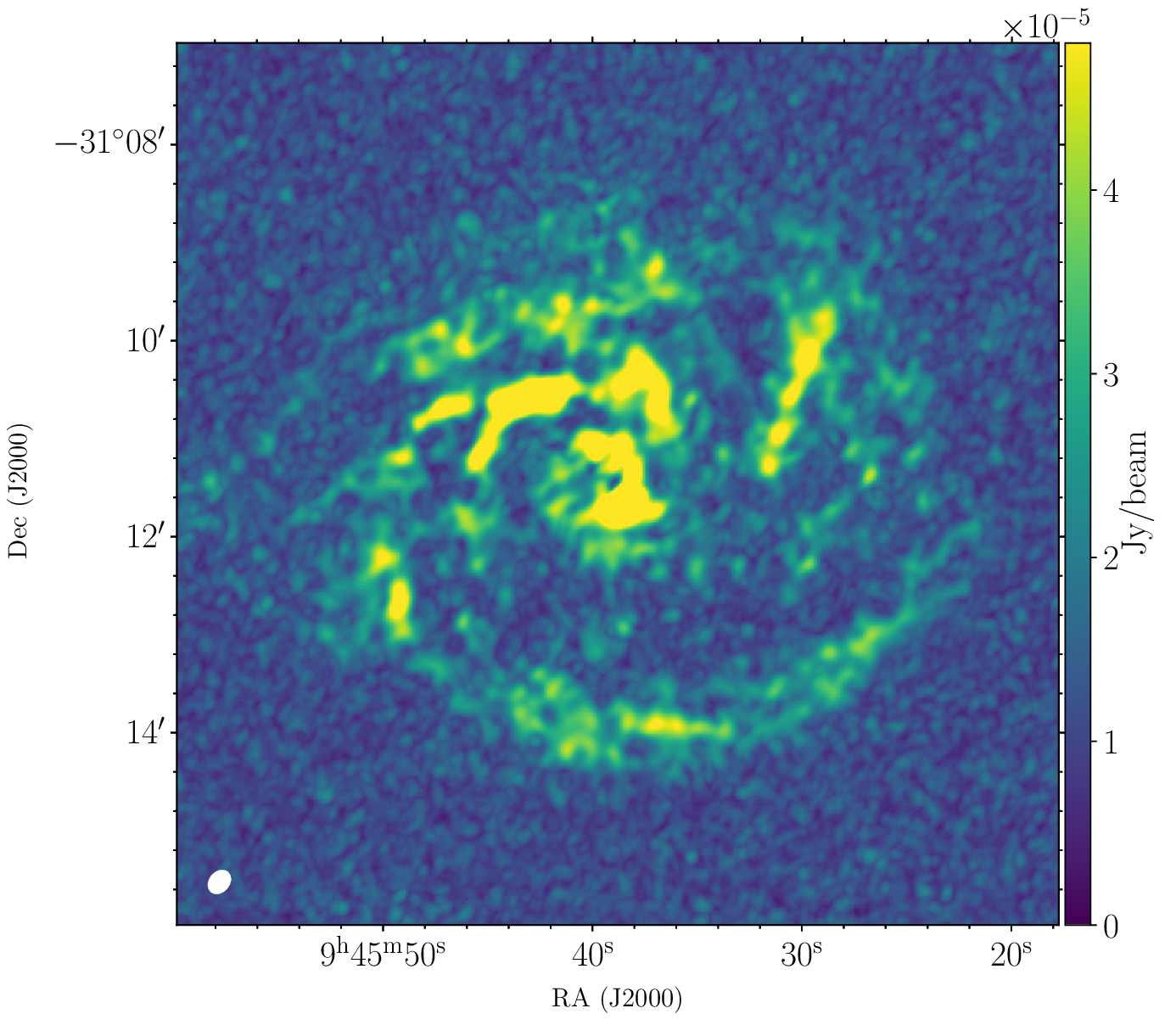}
        %\caption{Caption for the second figure}
        \label{fig:PI_rob0}
    \end{minipage}
    \caption{{\bf Left:} Polarized intensity of NGC\,2997 in color scale using robust parameter of 2 for the weighting and contours, corresponding to the total radio intensity. The scales are in Jy\,beam$^{-1}$. The contour levels are (3, 5, 8, 16, 32, and 50)\,$\times\,15.5\,\mu\rm{Jy\,beam}^{-1}$ at a resolution of $32\arcsec\times26\arcsec$. The beam size is shown at the bottom left corner of the image. The intrinsic polarization angles are rotated by 90\degr\ to show the orientations of the ordered magnetic field on the plane of the sky. {\bf Right:} Polarized intensity of NGC\,2997 in color scale using robust 0 weighting at a resolution of $15.2\arcsec\times11.6{\arcsec}$.}
    \label{fig:PI_PI_rob0}
\end{figure*}

Figure \ref{fig:total_I} shows the total radio intensity of NGC\,2997 at different angular resolutions. The image in the left panel was generated using a robust parameter of 0. Notably, the most prominent emission emanates from the galaxy's core. At this specific resolution of $12.8\arcsec\times9.8{\arcsec}$, a root mean square (rms) of $11\,\mu\rm{Jy\,beam}^{-1}$ is achieved. However, the central region of the galaxy cannot be resolved at this resolution.

From the map based on a robust parameter of 0, we measured a total flux density, integrated over the whole galaxy out to a radius of 250\arcsec, of $0.227\pm0.007\,\rm{Jy}$ (assuming a calibration error of 3\%).
At 2.4\,GHz with ATCA, this galaxy exhibits a continuum flux density of $0.16\pm0.01\,\rm{Jy}$ which is a lower limit due to missing large-scale flux density of the ATCA data \citep{Han_1999}. This is expected, as MeerKAT data with its dense core suffers less from missing extended emission problems.

Figure \ref{fig:total_I} (left panel) reveals two strong spiral arms emerging from the inner ring, plus two fainter arms splitting off the main ones. Many intensity peaks correspond to star-forming regions. Additionally, we observe a void of radio flux in the northwestern arm, disrupting the continuity of the spiral arm pattern. This phenomenon is consistent with observations at other wavelengths, such as optical observations. Figure \ref{fig:total_I} (right panel) shows the central region in more detail. The resolution of this image of $4.6^{\prime\prime}\times3.7^{\prime\prime}$ was achieved by using a robust parameter of -1. The rms of this map is $1.5\,\mu\rm{Jy\,beam}^{-1}$. Lower resolution studies described this region as a ring \citep{Kodilkar_2011}, but  the high resolution of these MeerKAT observations reveals an open ring, suggesting this could be the starting point of one of the spiral arms in this galaxy.

Figure \ref{fig:flux_literature} shows the comparison of different studies with the current MeerKAT observations. A linear fit with measurements from other studies gives a spectral index of the total flux densities of $-0.85\pm 0.04$. For this fit,
a number of values from the literature (Table~\ref{flux_literature_table}) were discarded because they were based on either short snapshot observations or displayed overly coarse $uv$ coverage; in such  cases, the extended emission of the galaxy could not be fully detected.

The inset in Figure \ref{fig:flux_literature} shows the total intensity flux density of individual spectral windows at the robust 0 weighting. The slope of these data points is $-1.14\pm 0.01$. The steepness compared to the overall spectrum is an artifact due to the variation in the innermost $uv$ coverage across the frequency band, but does not affect the further analysis of this study.

\begin{figure}
    \centering
        \includegraphics[width=\columnwidth]{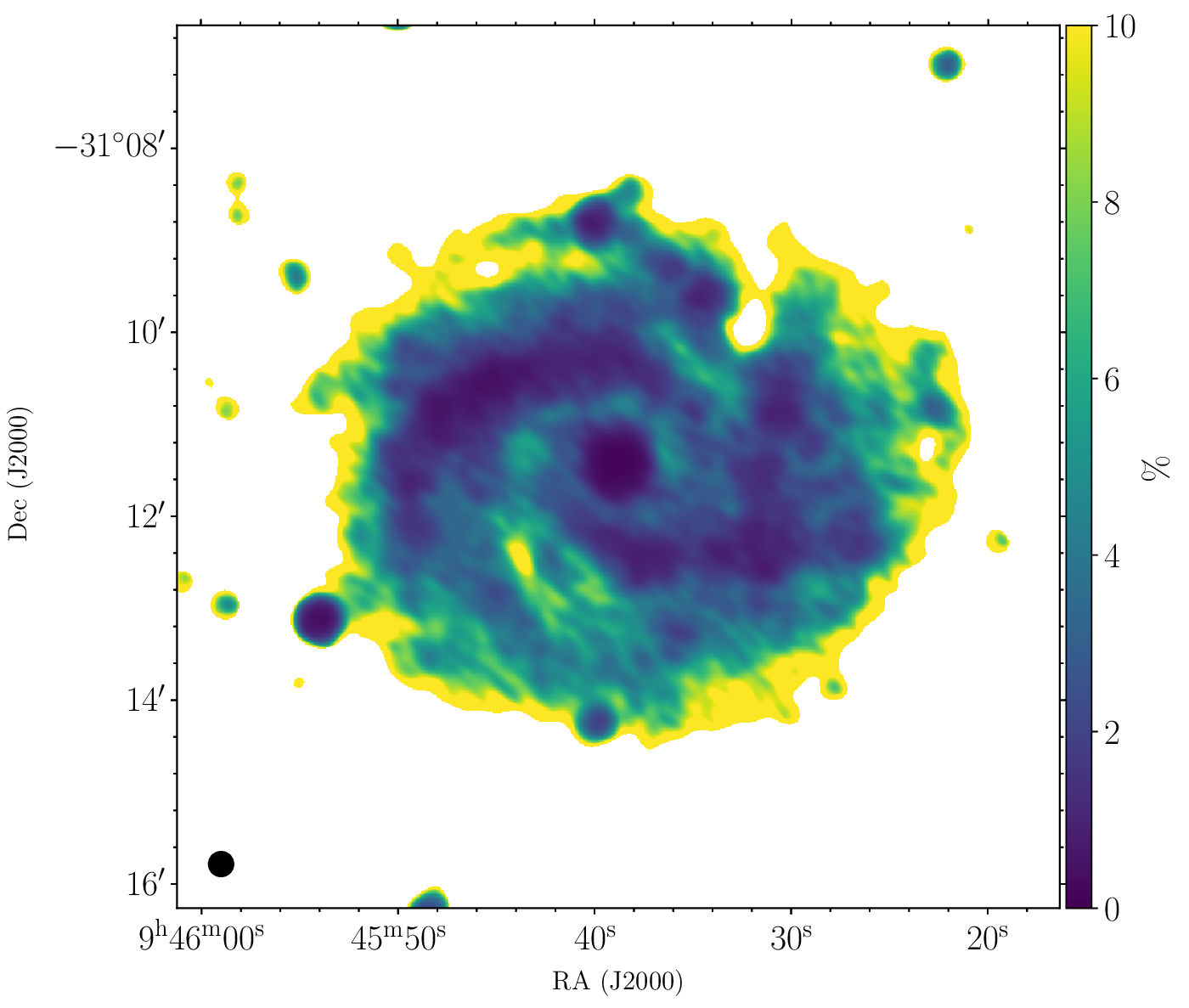}    
    \caption{Degree of polarization map of NGC\,2997. The color scale indicates the percentage of polarized signal. The beam size of $16\arcsec \times 16\arcsec$ is represented at the bottom left corner with a black circle. The thermal contribution of the total intensity data has not been removed. An estimate of the thermal emission with the help of an IR image at $24\,\mu$m would be possible, but is beyond the scope of this paper.}
    \label{fig:degree_of_pol}
\end{figure}

\begin{figure*}
    \centering
    \begin{minipage}{\columnwidth}
        \centering
        \includegraphics[width=\columnwidth]{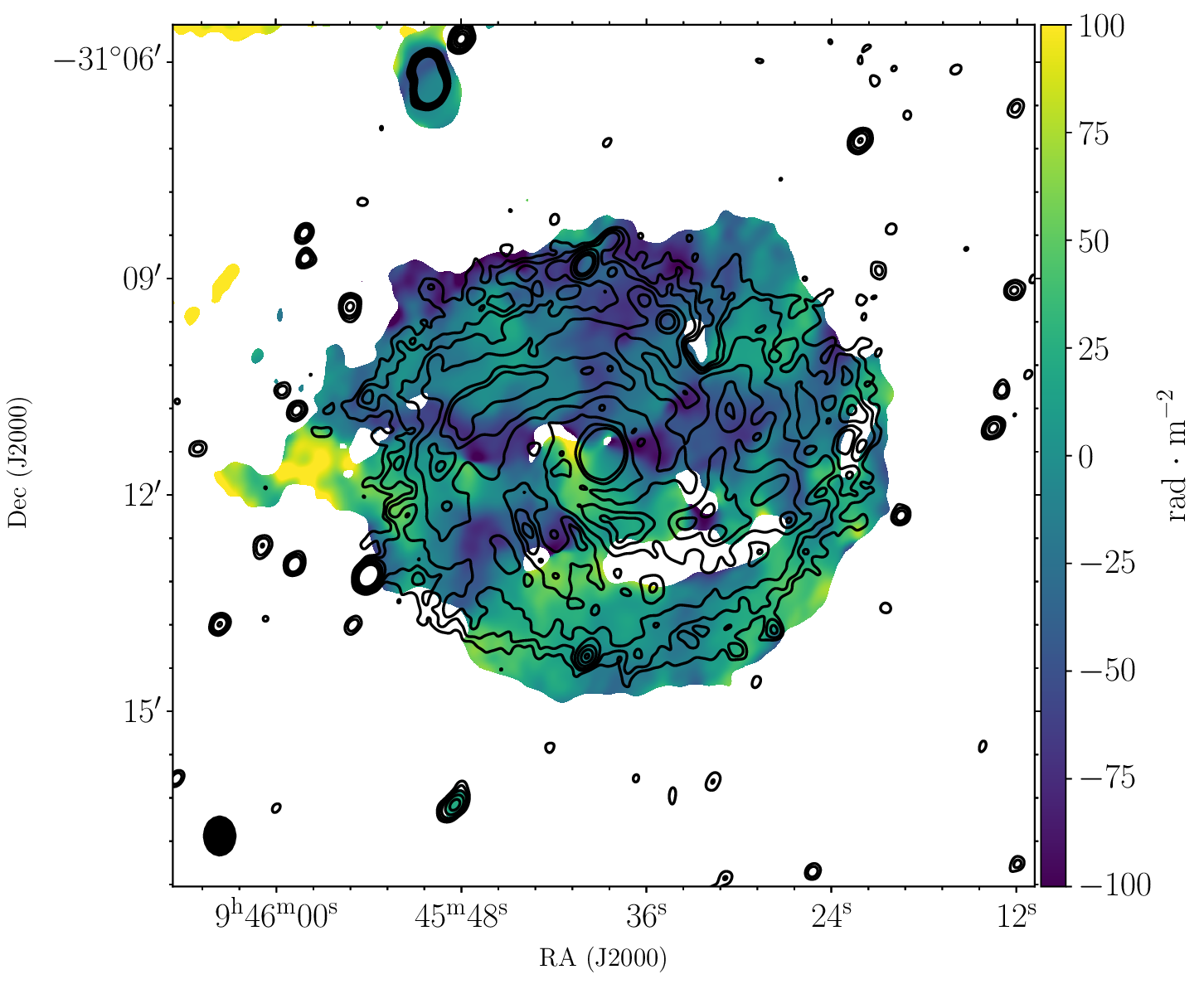}
    \end{minipage}
    \begin{minipage}{\columnwidth}
        \centering
        \includegraphics[width=\columnwidth]{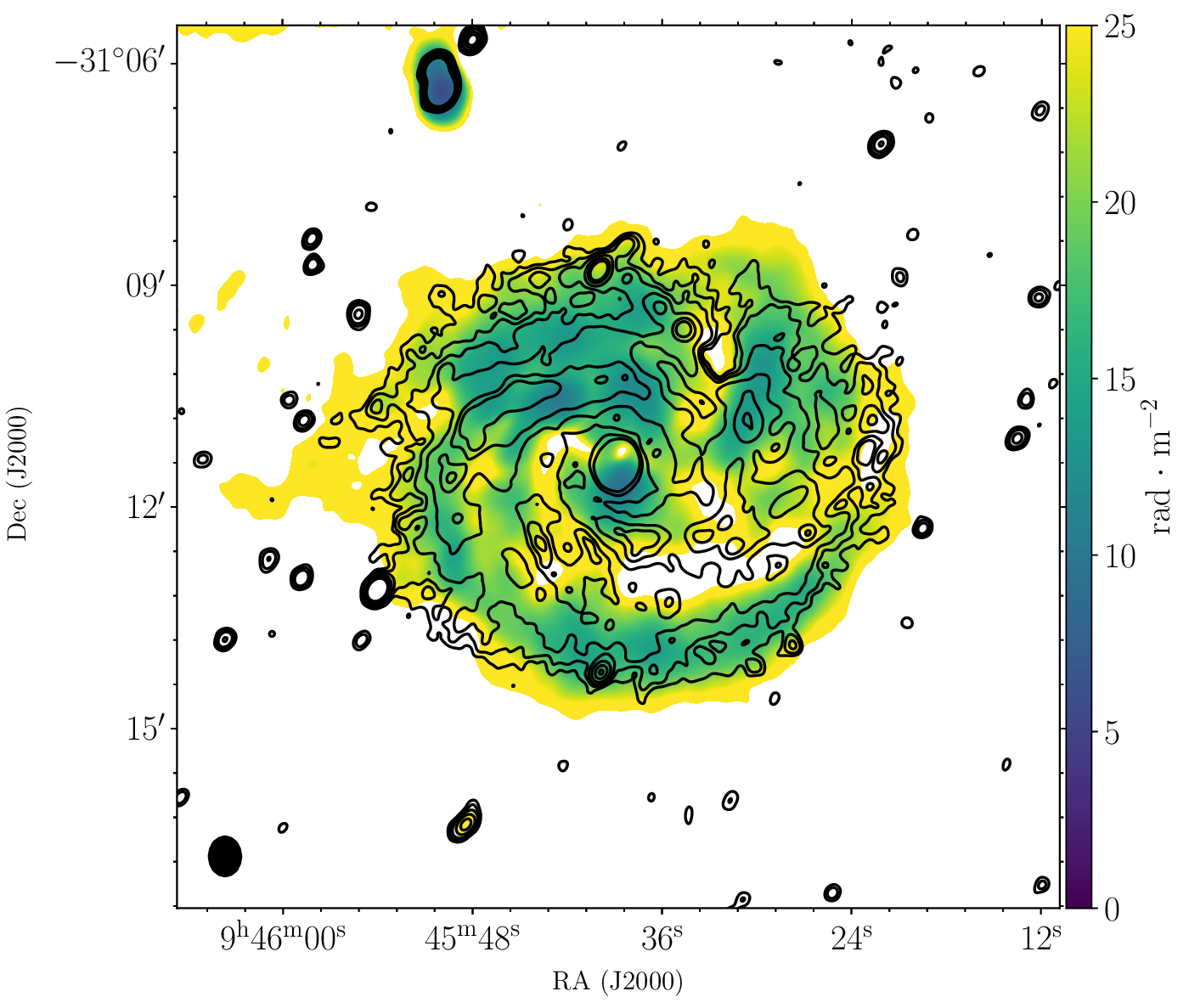}
    \end{minipage}
    \caption{{\bf Left:} Total radio intensity contours on top of a color scale map of the RMs after applying RM synthesis. The scales are in $\rm{rad\,m}^{-2}$. The contour levels are (3, 5, 8, 16, 32, and 50)\,$\times\,15.5\,\mu\rm{Jy\,beam}^{-1}$ at a resolution of $32\arcsec\times26\arcsec$. The foreground RM of our Galaxy has not been subtracted.
    {\bf Right:} Same total intensity contours as in the left panel on top of the RM error map. The beam size is shown at the bottom left corner of each image. A cut-off of $5\times$ rms noise is applied to all these maps. Regions close to the minor axis of the disk are excluded due to unrealistic RM values.}
    \label{fig:RM}
\end{figure*}

\begin{figure}
    \centering
        \includegraphics[width=\columnwidth]{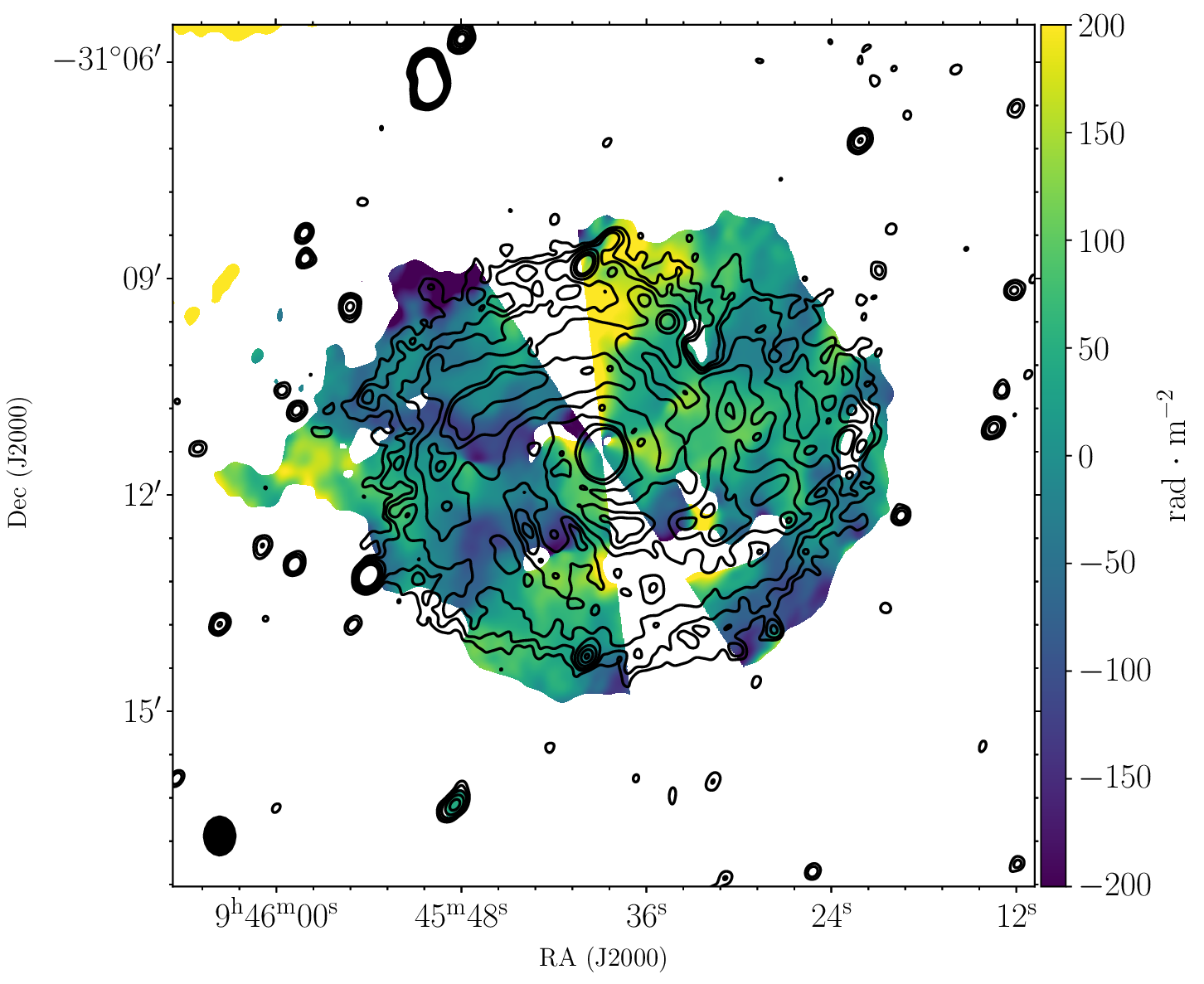}
    \caption{Same total intensity contours as in Figure \ref{fig:RM} on top of the RMs of the inclination-modified map described in in Section \ref{section:inclination}.}
    \label{fig:inclination}
\end{figure}

\subsection{Polarized emission}

The polarized emission in this galaxy is widespread across nearly the entire disk.
Individual regions of low polarization are aligned with the spiral arms, which is consistent with what it is observed in total radio intensity (see Figure \ref{fig:PI_PI_rob0} for a comparison between the left panel showing the map with a robust 2 weighting and the right panel showing the map with robust 0 weighting). The figure depicts total intensity contours combined with a color scale of polarization intensity. Additionally, the intrinsic polarization angles, rotated by $90\degr$ to reveal the orientations of the ordered magnetic field, are overlaid in Figures \ref{fig:PI_PI_rob0} and \ref{fig:optical_radio}.
An immediate observation from this map is the tight alignment of magnetic field orientations along the spiral pattern. Notably, the southern spiral arm exhibits an elongated area of polarization situated in the southern inter-arm region. The peak flux density in this region is $1.7\times 10^{-4}\,\rm{Jy \, beam}^{-1}$. Tracing its starting point from the northern part near the nucleus, this main magnetic spiral arm extends up to $14\arcmin$ in longitude, enveloping almost the entire galaxy. Given the galaxy's distance of 14.8\,Mpc, this corresponds to an ordered magnetic field spanning approximately 20\,kpc in longitude. The brightest area in polarized emission is located near the center of the galaxy where the peak flux density is $3.9\times 10^{-4}\,\rm{Jy \, beam}^{-1}$.

A map of the degree of polarization is shown in Figure \ref{fig:degree_of_pol}. The degree of polarization was computed for pixels with total intensity flux greater than five times the rms of the total intensity map of the left panel in Figure \ref{fig:total_I}.
The degree of polarization is smallest (below 3\%) in the central region and the spiral arms, reaches up to 10\% in the interarm regions, and increases beyond 10\% towards the outer regions. This behavior is similar to that observed in several other spiral galaxies \citep{Beck_2015} and indicates that the thermal contributions decreases and/or that the magnetic field becomes more ordered at larger radii, due to the decrease of star-formation activity.

Towards the eastern part of the disk we detect what seems to be the extension of another spiral arm. This structure is much fainter than the main magnetic spiral arm, but extends out of the disk and re-shapes the overall circular structure of the galaxy on the eastern part. This could be associated with the second prominent spiral arm that we detect in total intensity,  which is also observed at other frequencies. It might also indicate a past interaction with a neighboring galaxy, but this seems less probable given  the next large companion is more than 100\,kpc away.

The pitch angles between the optical arms and the magnetic arms are much larger in the northern and western parts of the disk compared to the southern and eastern parts (Figures~\ref{fig:sector_pitch_angle} and \ref{fig:optical_radio}). This could mean that the spiral-arm pattern observed in stars and gas does not fully control the magnetic pattern, but other agents could be at play, such as interactions with the circum-galactic medium or outflows from the disk \citep[e.g.,][]{Mulcahy_2017,Kierdorf_2020}.

\subsection{Rotation measures (RMs)}
\label{RM_analysis}

\begin{figure*}
    \centering
    \begin{minipage}{0.5\textwidth}
        \centering
        \includegraphics[width=1.0\textwidth]{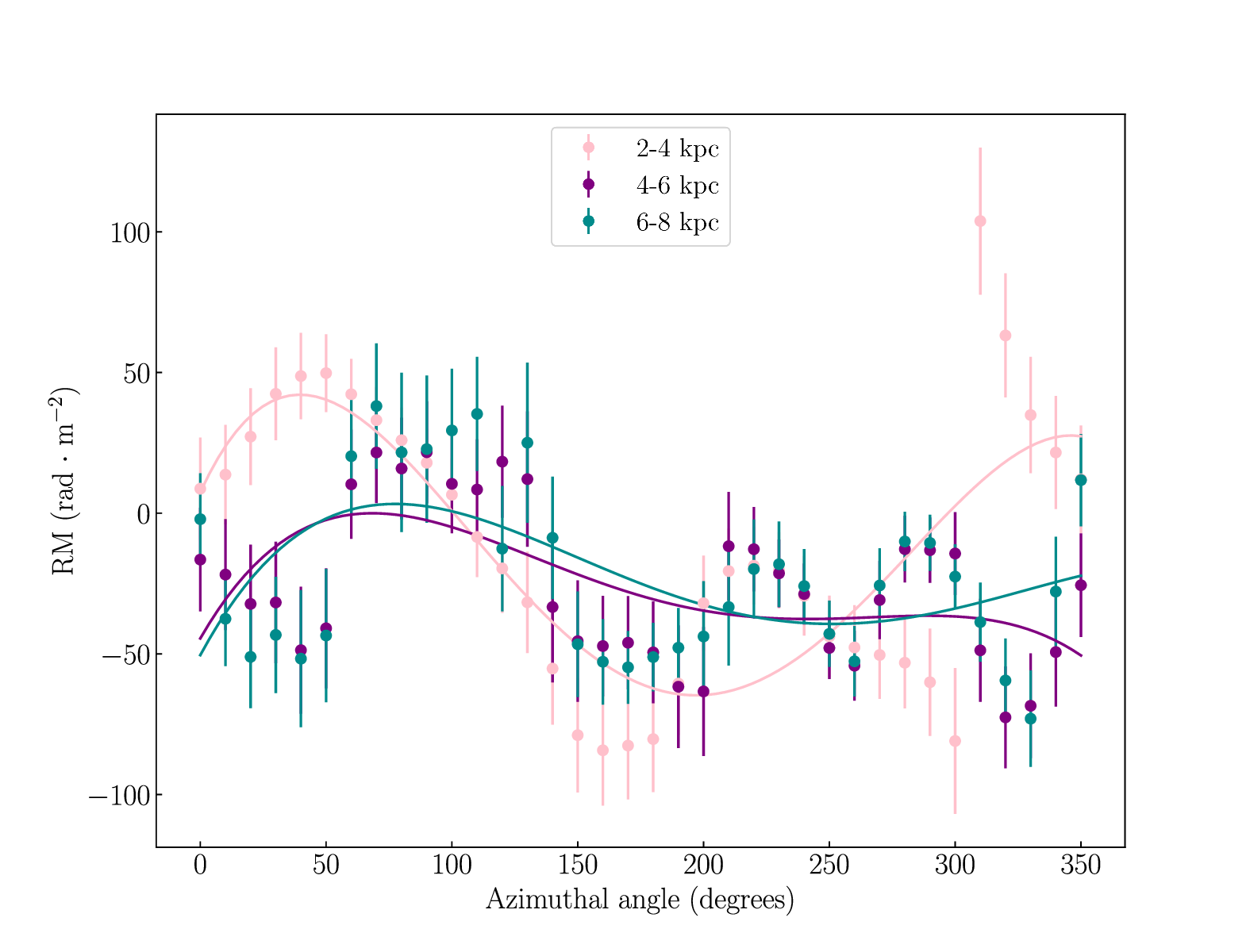}
    \end{minipage}\hfill
    \begin{minipage}{0.5\textwidth}
        \centering
        \includegraphics[width=\textwidth]{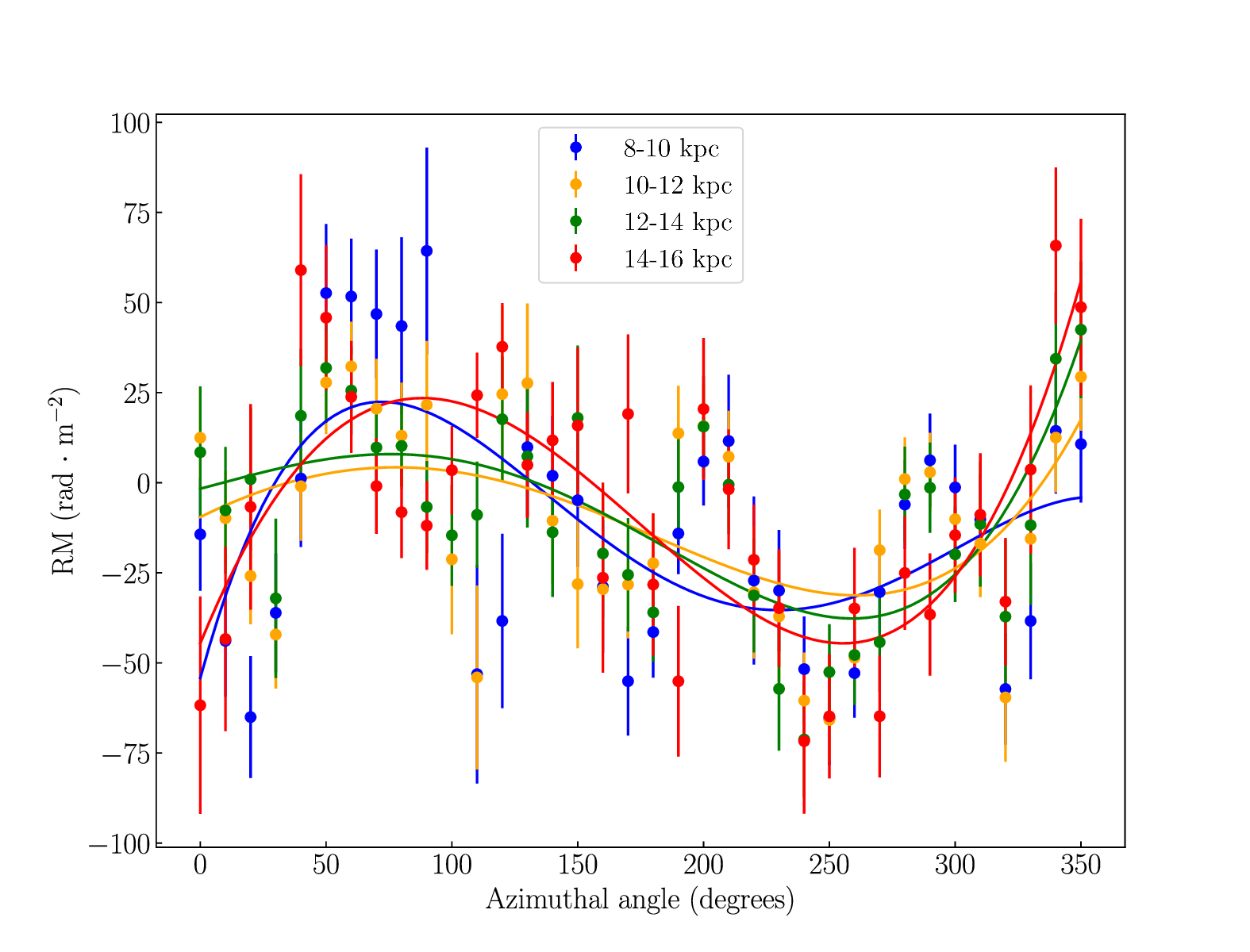}
    \end{minipage}

    \caption{Sector analysis of RMs in different rings, as described in Section \ref{RM_analysis}. Each color represents a different ring. Data points and estimated error bars are shown with the same color in each ring. Solid lines show fourth-order polynomial functions. The first ring starts between 2\,kpc and 4\,kpc from the center in the plane of the galaxy, and subsequent rings are placed every\,2 kpc. The last ring is placed between 14\,kpc and 16\,kpc where the polarization emission reaches three times the signal-to-noise ratio (S/N) value. Each ring is divided into multiple sectors separated by 10\degr\ in azimuthal angle. Sectors start at 0\degr\ on the south-eastern major axis of the galaxy on the plane of the sky and increase counter-clockwise. The plot was divided for a better visualization: The left panel shows rings from 2 to 8 kpc and the right panel shows rings from 8 to 16 kpc.}
    \label{fig:RM_sector}
\end{figure*}

\begin{figure}
        \begin{minipage}{0.5\textwidth}
        \centering
        \includegraphics[width=\textwidth]{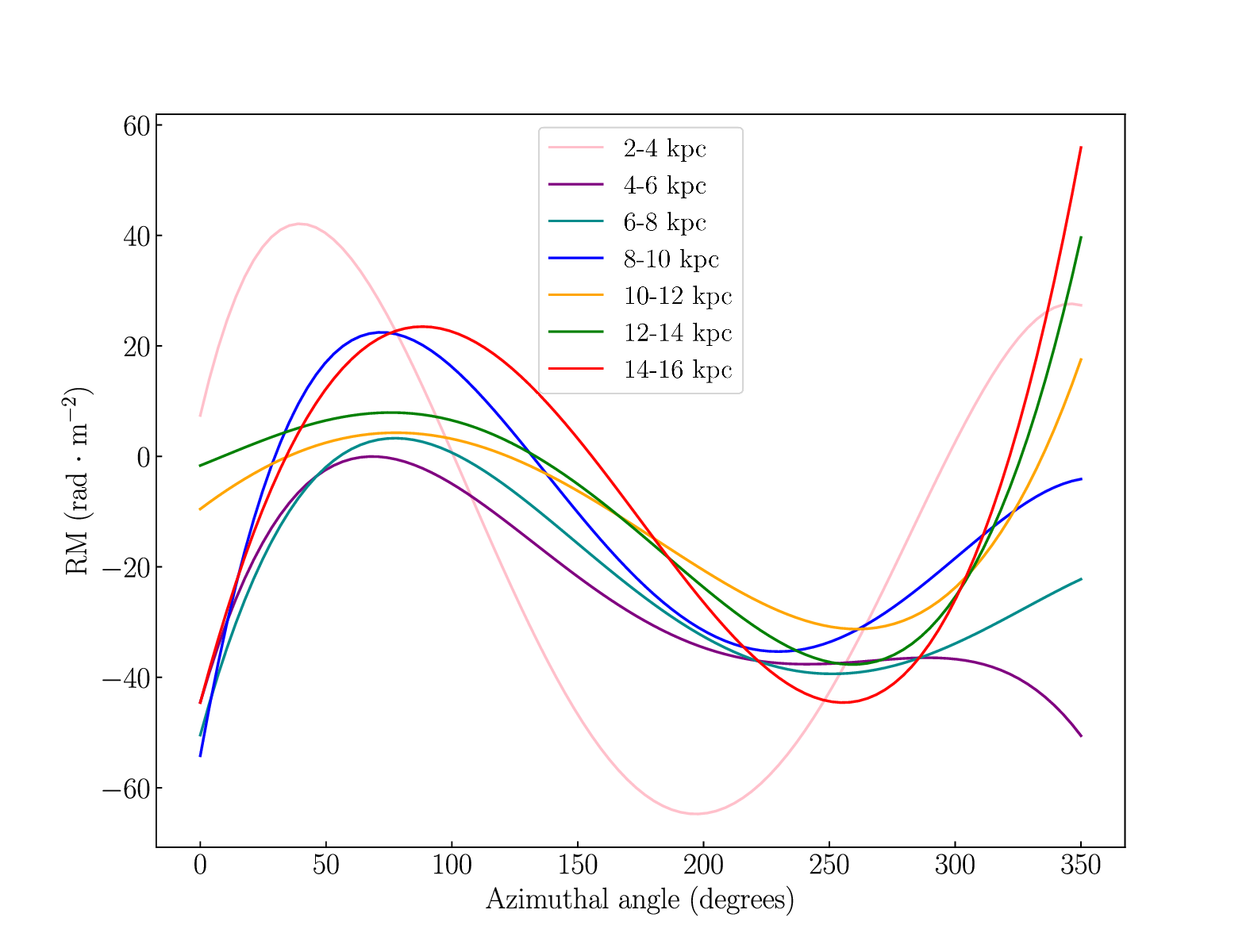}
    \end{minipage}
    \caption{Same fitting functions of Figure \ref{fig:RM_sector} corresponding to all the sectors used in this analysis are given for clarity.}
    \label{fig:RM_sector_fits}
\end{figure}

The RM synthesis analysis of our study at $32\arcsec\times26\arcsec$ resolution shows a complex pattern and large-scale gradients of RMs across the disk (Figure \ref{fig:RM}, left panel). For the RM analysis, we only considered pixels with total intensity flux greater than five times the rms of the total intensity map in Figure \ref{fig:total_I} (left panel).
To gain deeper insights into the large-scale structure of the regular magnetic field, we conducted a detailed analysis of the maps of RM and RM error across concentric rings in the plane of the galaxy, segmented into sectors along the azimuthal angle with a width of $10\degr$.

The average RM and the average RM error were computed from all pixels in a sector,
using weights proportional to the inverse of the RM error squared. The average RM error was divided by the square root of the number of beams in the relevant sector, giving the error of the average RM in the sector. The results of this analysis are illustrated in Figure~\ref{fig:RM_sector}. In this figure, the galaxy's disk is partitioned into seven rings in the plane of the galaxy, separated by 2\,kpc (similar to the spatial resolution at the galaxy's distance), highlighting similar azimuthal variations in the RM at all radial distances. Fourth-order polynomial functions are shown to represent the variations (see Figure~\ref{fig:RM_sector_fits}).
The real variation is more complex and calls for a more sophisticated
analysis (Section~\ref{section:Fourier}).

\begin{figure*}
    \centering
    \begin{minipage}{0.50\textwidth}
        \centering
        \includegraphics[width=\textwidth]{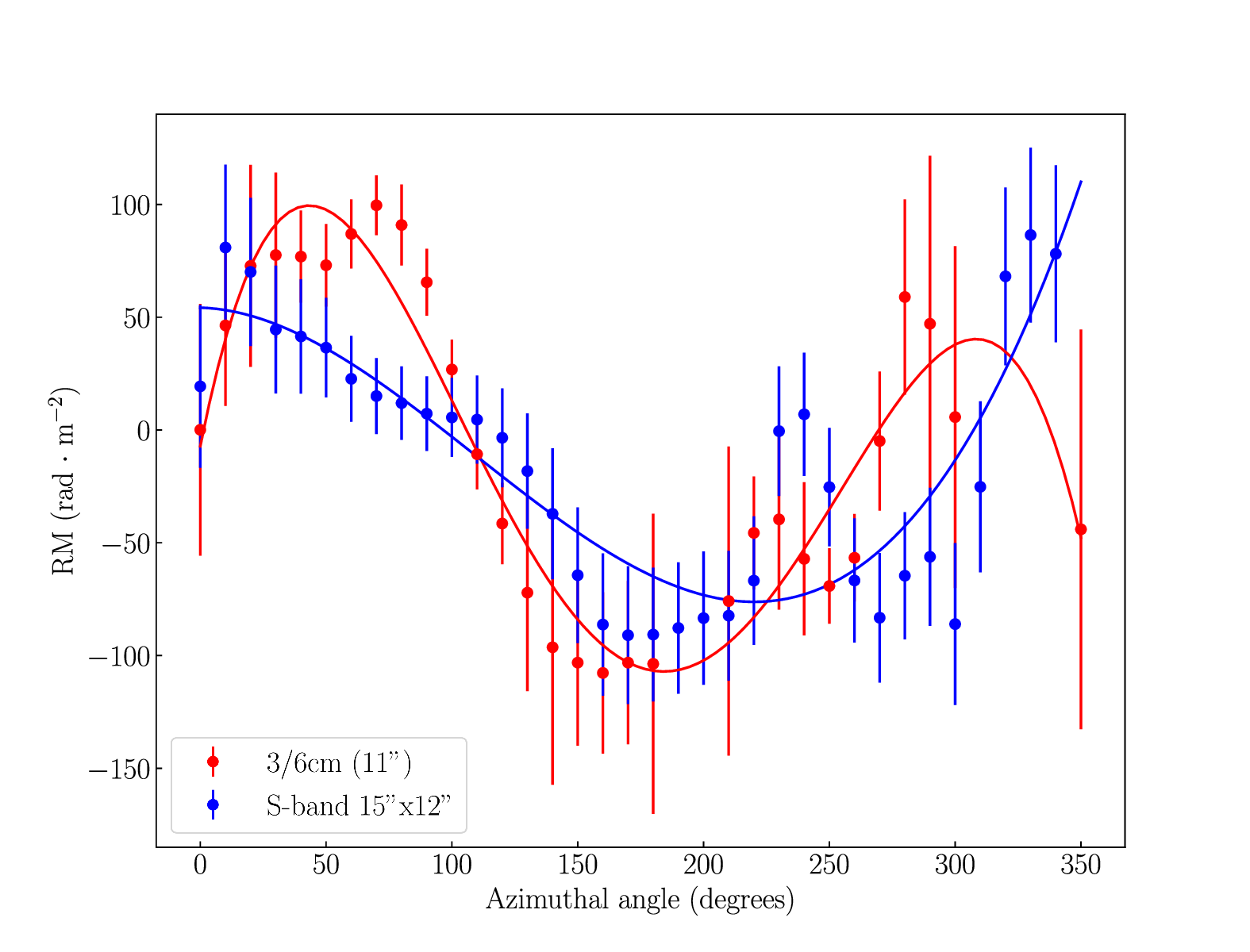}
    \end{minipage}\hfill
    \begin{minipage}{0.50\textwidth}
        \centering
        \includegraphics[width=\textwidth]{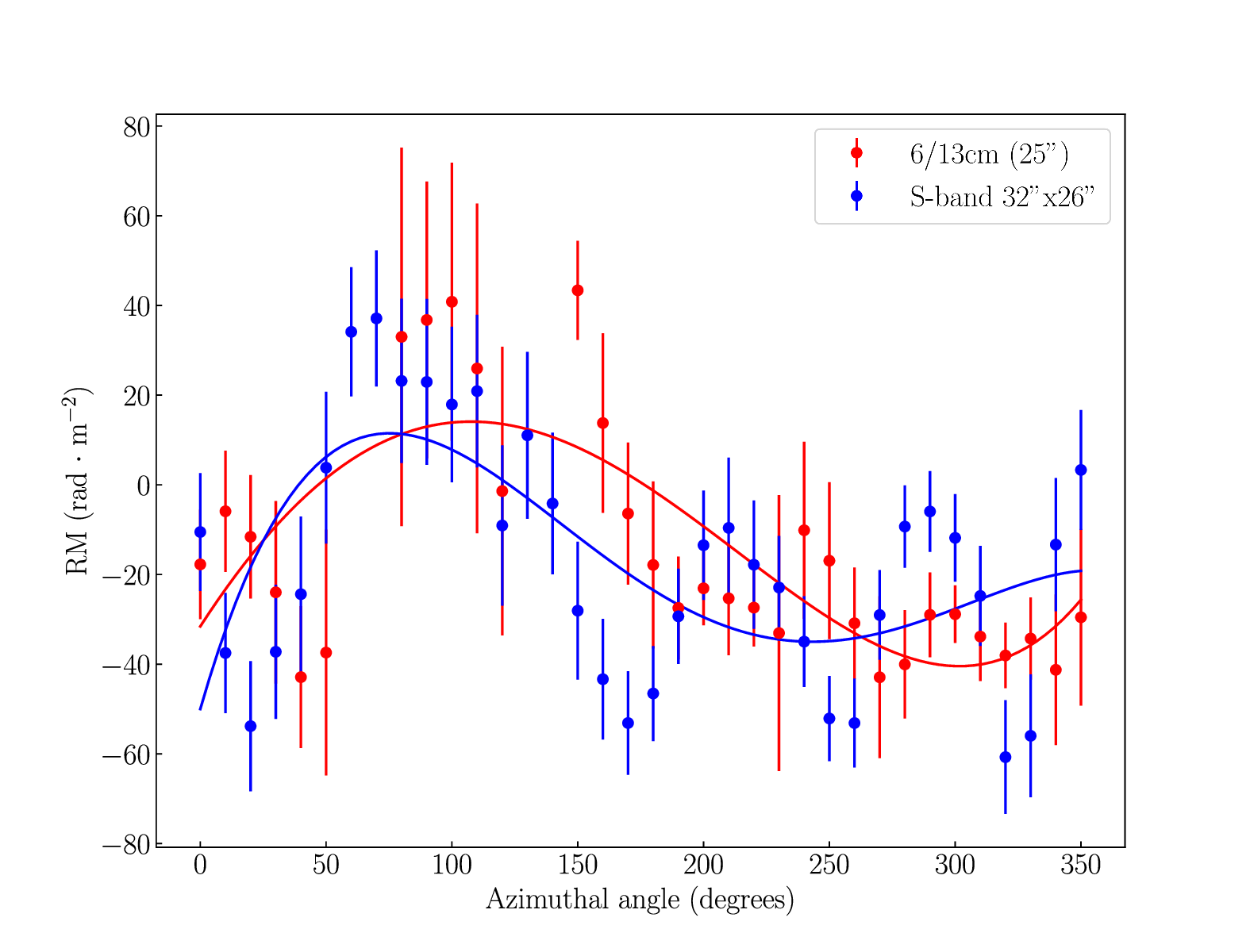}
    \end{minipage}
    \caption{{\bf Left:} RM variations along the azimuthal angle in the inner ring between 0.5\,kpc and 2\,kpc radius in the plane of the galaxy, based on the robust 0 data at $15\arcsec \times 12\arcsec$ resolution presented in this paper (blue curve), compared with the previous 6\,cm/13\,cm data at $11\arcsec$ resolution from \cite{Han_1999} (red curve). Data points are shown with the estimated error bars of each ring. Solid lines show fourth-order polynomial functions. {\bf Right:} RM variations along the azimuthal angle in the ring between 4 kpc and 10 kpc radius in the plane of the galaxy derived from the robust 2 data at $32\arcsec \times 26\arcsec$ resolution of this paper (blue curve) and from the previous 6\,cm/13\,cm data at $25\arcsec$ resolution from \cite{Han_1999} (red curve).}
    \label{fig:RM_sector_inner}
\end{figure*}

The angular resolution of $25\arcsec$ of the RM map between 2.4\,GHz and 4.86\,GHz in \citet{Han_1999} is comparable to the resolution of our robust 2 weighting data (Figure ~\ref{fig:RM}). Our RM variation is similar to that in \citet{Han_1999} for the same ring between 4\,kpc and 10\,kpc (Figure~\ref{fig:RM_sector_inner}).
The RMs of the current study are not affected by gaps at certain azimuthal angles due to weak signals, as in \citet{Han_1999}, and the RM errors are smaller.

\cite{Han_1999} discussed a possible reversal between the disk field in the radial range 4 -- 10\,kpc / azimuthal range $\sim 300\degr-50\degr$ (i.e. east -- south) with negative RMs (see Figure~\ref{fig:RM_sector_inner}, right panel) and the field in the inner region (radial range 0.5 -- 2.0\,kpc and the same azimuthal range) with positive RMs, obtained between 4.86\,GHz and 8.46\,GHz with $11\arcsec$ resolution.
To check this claim, we averaged the RMs based on our robust 0 data at $15\arcsec\times12\arcsec$ resolution in the same radial range of 0.5 -- 2.0\,kpc. We confirm the positive RMs in the azimuthal range $\sim 330\degr-50\degr$ with higher accuracy (Figure~\ref{fig:RM_sector_inner}, left panel).

Importantly, we did not find any evidence of a field reversal with respect to the disk, as claimed by \cite{Han_1999}. The RM variation in the first ring of disk of 2 -- 4\,kpc (Figure~\ref{fig:RM_sector}, left panel) is similar to that in the inner ring of 0.5 -- 2.0\,kpc (Figure~\ref{fig:RM_sector_inner}, left panel), without the need for a field reversal.
The RM variations in the following rings smoothly shift in phase. In conclusion, the RM variation in such a wide ring as the one used in Figure~\ref{fig:RM_sector_inner} (right panel) smears out the structure of the regular field; thus, it is not suited to measuring the field direction.

The average $\rm{RM}_{fg}$ from the foreground of our Galaxy was estimated using the weighted average of the offsets (zero lines) from all fits to the RM variations among all the rings, which yields a value of $\rm{RM_{fg}}= (14\pm17)\,\rm{rad\cdot m^{-2}}$.

\begin{figure}
        \centering
        \includegraphics[width=\columnwidth]{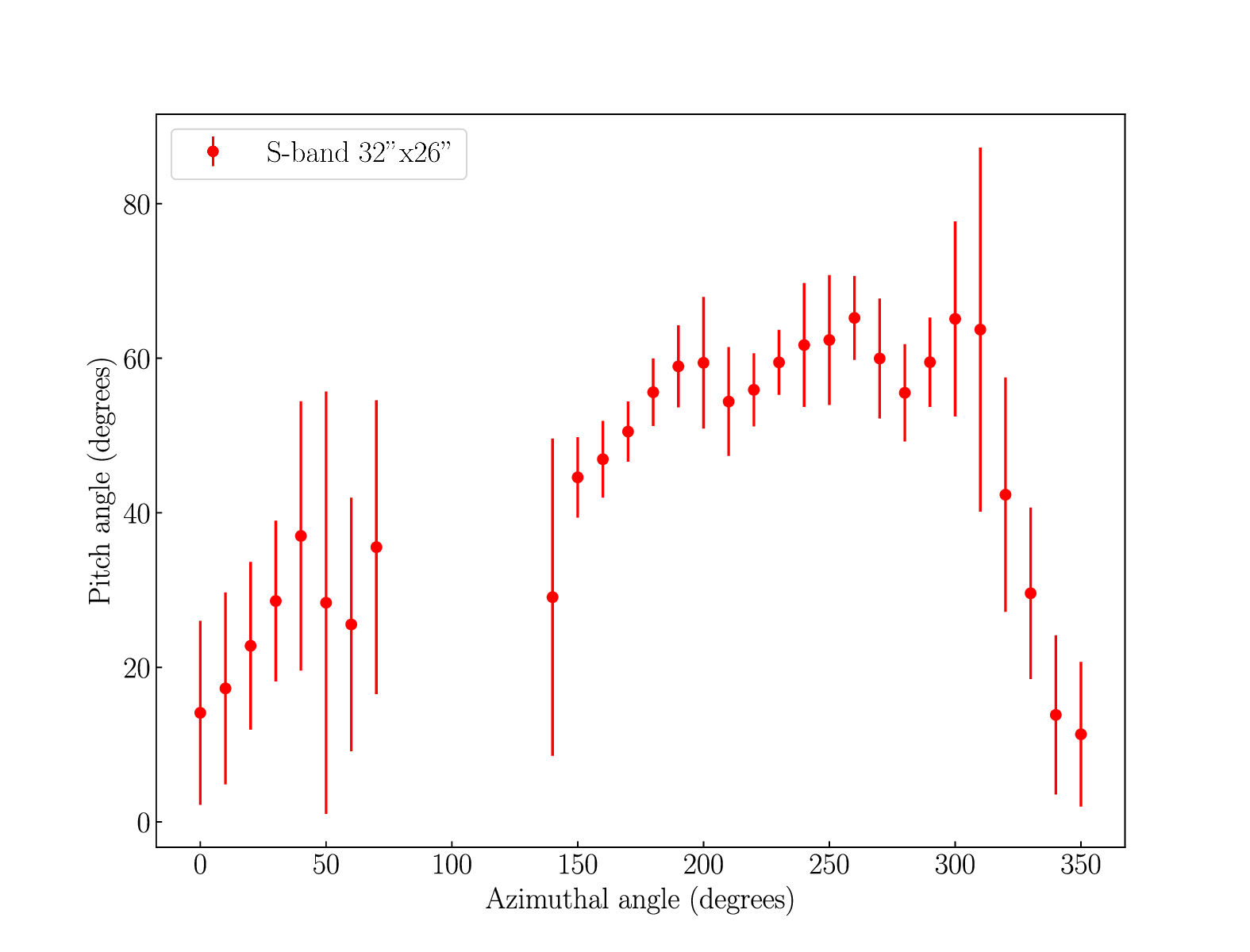}
        \caption{Variation in the spiral pitch angle of the intrinsic polarization angles along the azimuthal angle in the ring between 4\,kpc and 10\,kpc radius in the plane of the galaxy, derived from the robust 2 data at $32\arcsec \times 26\arcsec$ resolution.
        Sectors start at 0\degr\ on the south-eastern major axis of the galaxy on the plane of the sky and increase counter-clockwise.
        The errors were estimated from the standard deviations of $Q$ and $U$ values within each sector.}
        \label{fig:sector_pitch_angle}
\end{figure}

As a byproduct of computing the RM variation along the azimuthal angle, we measured the intrinsic pitch angle of the magnetic field, corrected for RM, in the ring between 4 kpc and 10 kpc radius, derived from the robust 2 data (Figure~\ref{fig:sector_pitch_angle}). The pitch angle is $\sim 10\degr - 30\degr$ in the
southern half (azimuthal range $330\degr - 140\degr$), but larger in the northern half.

\section{Analysis and discussion}
\label{sec:discussion}

\subsection{Geometric effect on RMs}
\label{section:inclination}

Overall, RMs are affected by the galaxy's inclination. For a mathematical description of this effect, we assume a simple regular field with an axisymmetric spiral pattern in the galaxy plane, consisting of azimuthal and radial components, with a constant pitch angle and constant strength, but without reversals,  out-of-plane components, or Faraday depolarization.

Two general types of Faraday depolarization have been described in the literature, namely, differential Faraday rotation in regular magnetic fields and internal Faraday dispersion in turbulent fields \citep{Sokoloff1998}. Faraday dispersion is dominant in galaxy disks \citep{Williams_2024}. With a RM dispersion due to turbulent fields of $\leq30\,\rm{rad\cdot m^{-2}}$ (computed from the RM map shown in Figure~\ref{fig:RM}, left panel), the depolarization factor in the S band is DP $\geq0.8$ (i.e., less than 20\% of the polarized intensity is depolarized). The model of an axisymmetric spiral field has been first introduced by \citet{Krause_1989} for the spiral galaxy IC\,342.

The axisymmetric field dominates in most spiral galaxies observed so far (see Table 5 in \citet{Beck_2019}). The axisymmetric field is regarded as the first azimuthal mode ($m=0$) generated by the large-scale dynamo \citep[e.g.][]{Beck_1996}.
In an axisymmetric spiral field, the observable RM is
\begin{equation}
\rm{RM} = \rm{RM}_{fg}\,+\,\rm{RM}_{0}\,\rm{cos}(\theta - \psi)\,\rm{sin}(i)
,\end{equation}
where $\rm{RM}_{fg}$ is the RM of the Galactic foreground,
$\theta$ is the azimuthal angle in the plane of the galaxy, measured counter-clockwise from the eastern major axis, $\psi$ is the spiral pitch angle in the plane of the galaxy, and $i$ is the galaxy's inclination.
RM$_{0}$ is the inherent RM computed as 
RM$_0$ = 0.81 n$_\mathrm{e}$ B$_\mathrm{reg}$ L,
where n$_\mathrm{e}$ and B$_\mathrm{reg}$ are the average electron density and the average regular field strength along the line of sight, L. In face-on view (0\degr\ inclination), this regular magnetic field is entirely parallel to the plane of the sky, so that RM becomes zero.

When the disk is tilted at some degree of inclination, the axisymmetric magnetic field exhibits components that are parallel to the line of sight across the entire disk, except for points close to the minor axis of the galaxy. It is noteworthy that the direction of the magnetic field along the line of sight is contingent upon the positions of the nearest and furthest sides of the minor axis of the disk, along with the azimuthal direction of the field across the disk. This underscores the intricate relationship between inclination, magnetic field orientation, and the resultant RMs in scenarios with inclined galaxies.

Based on the above assumptions, the RM map
from Figure~\ref{fig:RM} (left panel) is modified for geometry (inclination and azimuthal angle) as follows,
\begin{equation}
\rm{RM}_{0} = (\rm{RM} - \rm{RM}_{fg})\, / \,(\rm{cos}(\theta - \psi)\,\rm{sin}(i))\, ,
\end{equation}

\noindent
resulting in Figure~\ref{fig:inclination}. Regions in close proximity to the minor axis of the disk, where the field component along the line of sight approaches zero, are excluded due to unrealistic RM values. This limit was set so that $-0.2 < \rm{RM}/\rm{RM}_{\rm{modified}} < 0.2$.

The RM modification yields two notable effects. Firstly, the range of RM values expands, now covering approximately $\pm 100\,\rm{rad\,m}^{-2}$, larger than in the original RM map. Secondly, a change in the sign of RMs is observed in the northern part of the disk (west of the minor axis). This inversion is a consequence of the modification, indicating that, post-modified, this region of the disk is now "approaching" the plane of the sky.

If NGC\,2997 were hosting a purely axisymmetric field, Figure~\ref{fig:inclination} would reveal constant $\rm{RM}_{0}$ values. Instead, the significant variations indicate deviations from a simple axisymmetric field, such as a varying pitch angle (Figure~\ref{fig:sector_pitch_angle}) and field strength, along with field reversals, higher dynamo modes, local field patterns, Parker instabilities, or other types of vertical fields.

Furthermore, the smooth phase shifts between adjacent rings (Figure~\ref{fig:RM_sector}, left panel) are inconsistent with a purely axisymmetric field. For higher azimuthal modes, the regular field strength is not constant and the position angles of the maximum field strength vary with radius. The search for higher modes is discussed in Section~\ref{section:Fourier}.

The determination of the detailed RM pattern is also contingent upon the shape of the halo field, which remains unknown due to the galaxy's position in the sky. The halo field of edge-on galaxies is not parallel to the galactic plane and often reveals an X-shape pattern in field orientation, while large-scale patterns in RM have not been observed thus far \citep{Krause_2020}.
A more realistic scenario would involve alternating field directions, similar to the case observed in NGC\,4631 \citep{Mora_2019}, indicative of Parker instabilities \citep{Tharakkal_2023}. In such a vertical field configuration, RM fluctuations would overlay onto the RM derived from the disk field, manifested as small-scale RM variations along the azimuth.

\subsection{Fourier analysis of the RM variations}
\label{section:Fourier}

Patterns of magnetic spiral arms can be seen in the RM maps of face-on galaxies and were measured by studying azimuthal RM profiles across the disk \citep[e.g.,][]{Fletcher_2004, Beck_2019}. In dynamo theory, the basic (axisymmetric) azimuthal dynamo mode is predicted to reveal the largest amplitude, while the amplitudes of higher modes depend on the number of spiral arms characterized by an enhanced star formation rate and turbulence \citep{Chamandy_2013a, Chamandy_2013, Chamandy_2014}.

\begin{figure}
\centering
        \includegraphics[width=\columnwidth]{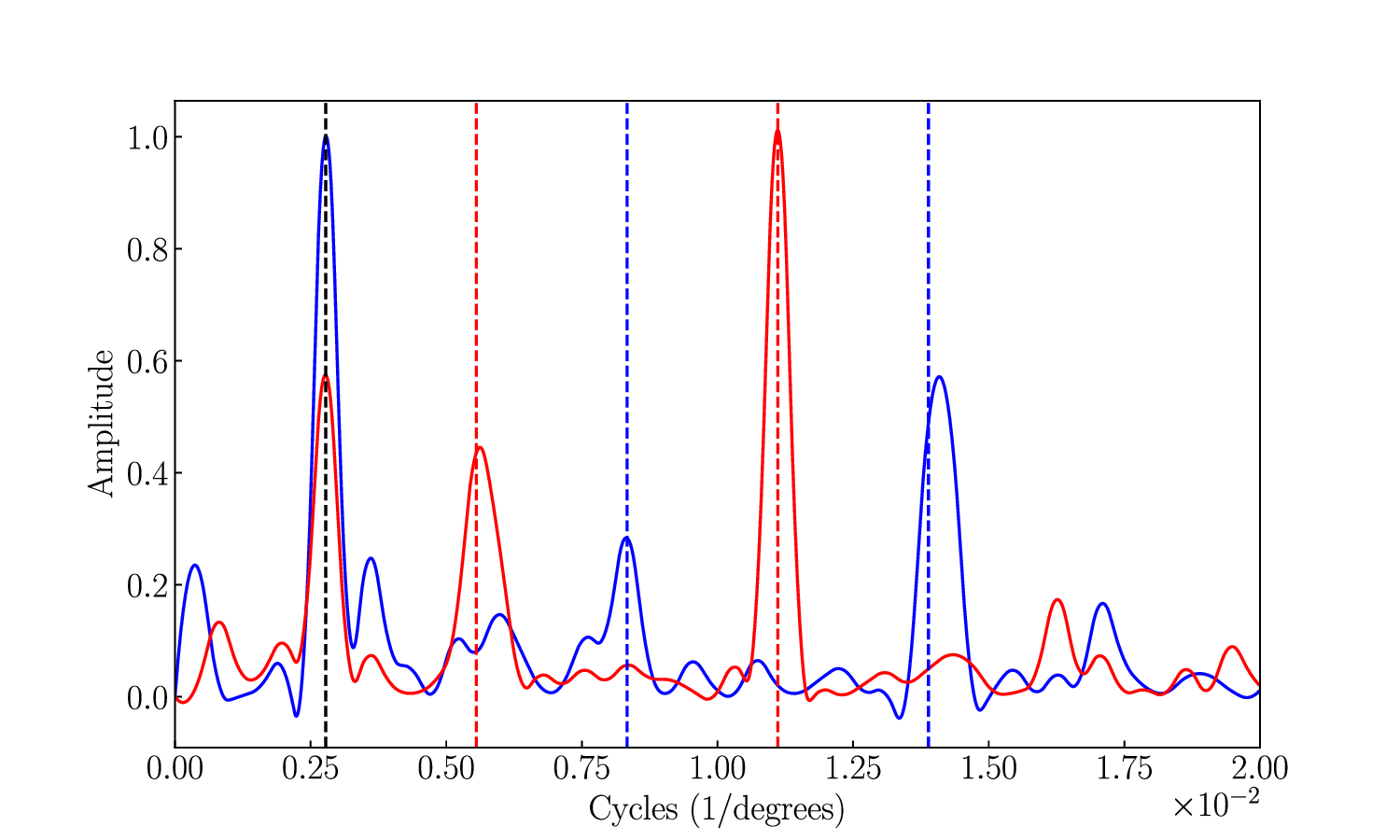}
        \caption{Results of the Fourier analysis on the RM data of the different sectors presented in Figure~\ref{fig:RM_sector}. The blue curve represents the FT of the original RM data. The red curve represents the FT of the geometrically modified RM data as described in Section \ref{section:inclination}. Each dashed line shows the exact position of integers divided by 360 degrees. The black dashed line shows 1/360, the red dashed line shows 2/360 and 4/360, and the blue dashed line shows 3/360 and 5/360.}
        \label{fig:FFT}
\end{figure}

\begin{table*}
\caption{Fourier components detected in the original RM map.}
    \centering
    \begin{tabular}{c|c|c|c|c}
                \toprule
       Peak  & Cycles (1/degrees) & Relative amplitude & Phase w.r.t. peak 1 & Dynamo mode\\
                \midrule
        1 & $(2.7874\pm0.199) \times 10^{-3}$& 1 & 1 & 0\\
        2 & $(8.2427\pm0.364) \times 10^{-3}$ & $0.24\pm0.05$ & $2.96\pm0.25$ & 2\\
        3 & $(14.0927\pm0.281) \times 10^{-3}$ & $0.60\pm0.05$& $5.06\pm0.38$ & 4\\
             \bottomrule
    \end{tabular}
    
\label{tab:FFT1}
\end{table*}

\begin{table*}
\caption{Fourier components detected in the geometrically modified RM map.}
    \centering
    \begin{tabular}{c|c|c|c|c}
                \toprule
       Peak  & Cycles (1/degrees) & Relative amplitude & Phase w.r.t. peak 1 & Dynamo mode\\
                \midrule
        1 & $(2.767\pm0.221) \times 10^{-3}$ & $0.55\pm0.01$& 1 & 0\\
        2 & $(5.650\pm0.326) \times 10^{-3}$ & $0.43\pm0.05$& $2.04\pm0.20$ & 2\\
        3 & $(11.108\pm0.206) \times 10^{-3}$ & 1& $4.01\pm0.33$ & 4\\
                \midrule
    \end{tabular}
 
\label{tab:FFT2}
\end{table*}

For the case of NGC\,2997, we performed a fast Fourier transform (FFT) to describe the RM profiles shown in Figure~\ref{fig:RM_sector}. This analysis gives hints of the different modes that make up the magnetic field of a spiral galaxy. Three field components in cylindrical coordinates (radial, azimuthal, and vertical) for three azimuthal modes (m = 0, 1, and 2) were given in Eqs.~(2) of \cite{Fletcher_2004}. Azimuthal modes for the vertical field $\rm{B_z}$ are not predicted by dynamo theory for thin galaxy disks and are disregarded here, so that we were left with two equations. As we could not measure the phases $\beta_i$ of the azimuthal variations with our Fourier analysis, these parameters were also disregarded. The pitch angles, $\psi_i$, of the modes, $i$, affect the amplitudes of $\rm{B_r}$ and $\rm{B_{\theta}}$. For a zero pitch angle, the radial field component $\rm{B_r}$ vanishes completely. Small pitch angles indicate a dominating azimuthal component $\rm{B_{\theta}}$. The Fourier analysis allowed us to measure the amplitudes $\rm{B_i}$ of the modes, somewhat affected by the pitch angles $\psi_i$. The transformation from $\rm{B_r}$, $\rm{B_{\theta}}$, and $\rm{B_z}$ into the parallel field component (giving RM) was given in Eq.~(A2) of \citet{Berkhuijsen_1997}. The second line of this equation shows that an additional $\rm{cos}(\theta)$ dependence of all modes of $\rm{B_{\theta}}$ is caused by the inclination.

The blue curve in Figure~\ref{fig:FFT} shows the result of the FF on the original RM variations (Figure~\ref{fig:RM}, left panel). The analysis identifies different peaks that extend across the phase space of the Fourier Transform. For this study, we focused on the three largest peaks in each scenario. These peaks are selected to be at least 3$\cdot\sigma$ respect to the noise level $\sigma$ of the FT,
which was estimated to be 3.80 (0.07 in normalized units). The noise level was determined by fitting a Gaussian to the histogram of amplitudes. For comparative purposes, the amplitudes of both curves have been normalized. The first (main) peak, located around $0.28 \times 10^{-2} \,(1/\text{degrees})$, corresponds to dynamo mode 0, which recurs after every cycle of a ring (i.e., every 360 degrees).

Table~\ref{tab:FFT1} gives the positions and amplitudes of the three largest peaks with respect to the first peak. A Gaussian fit was performed to find the position and the error estimate for each peak.
The second and third largest peaks correspond to higher dynamo modes.
Their positions are almost exactly integer multiples (3 and 5) of the position of the first peak.

When applying the FFT to the geometrically modified RMs (red curve in Figure~\ref{fig:FFT}), the position of the first peak remains, while the second and third  peaks shift to positions that are still at almost perfect integer multiples (i.e., 2 and 4) of the position of the first peak (Table~\ref{tab:FFT2}).
Peaks 2 and 3 increase in amplitude after geometric modification. The first peak decreases in amplitude, while  this peak is expected to disappear completely for a perfectly axisymmetric mode with an azimuthal RM variation that is entirely caused by the inclination of the galaxy.
The fact that we can still see the first peak after the geometric modification means that our assumption of perfect symmetry has not been fulfilled. The regular field strength and the spiral pitch angle vary with azimuthal angle, as made clear from Figures \ref{fig:PI_PI_rob0} and \ref{fig:sector_pitch_angle}.

Mathematically, the geometric modification is expected to change
the azimuthal variations related to peaks 2 and 3 according to the following expressions:

\begin{equation}
    \frac{\rm{cos}3\theta}{\rm{cos}\theta} = 2\rm{cos}2\theta - 1
,\end{equation}
\begin{equation}
    \frac{\rm{cos}5\theta}{\rm{cos}\theta} = 2\rm{cos}4\theta - 2\rm{cos}2\theta + 1
,\end{equation}

\noindent where $\theta$ is the azimuthal angle of our analysis.
These expressions are also valid in case of a pitch angle $\psi$ (i.e., a phase shift in azimuthal angle), where $\theta$ is replaced by $\theta - \psi$.

The results presented in Tables~\ref{tab:FFT1} and \ref{tab:FFT2} are in excellent agreement with the expectation of distinct modes represented by integer multiples in cycles. We conclude that our Fourier analysis gives strong indication for the existence of three dynamo modes of an even order (i.e., modes 0, 2, and 4). Dynamo theory applied to a galaxy with two symmetric spiral arms predicts the excitation of modes 0, 2, and 4 with relative amplitudes 1, 0.36, and 0.06 \citep[see top left panel of Figure~7 in][]{Chamandy_2014}, to be compared with the relative amplitudes in Table~\ref{tab:FFT1}. The amplitude of mode 2 is consistent with theory, while mode 4 is about $10\times$ stronger than predicted, possibly enhanced by
the four-arm spiral pattern. Generally, a four-arm spiral is predicted to produce a four-arm magnetic spiral \citep[see Figure~14 in][]{Chamandy_2013a}.

\begin{figure*}
\centering
        \includegraphics[width=0.8\textwidth]{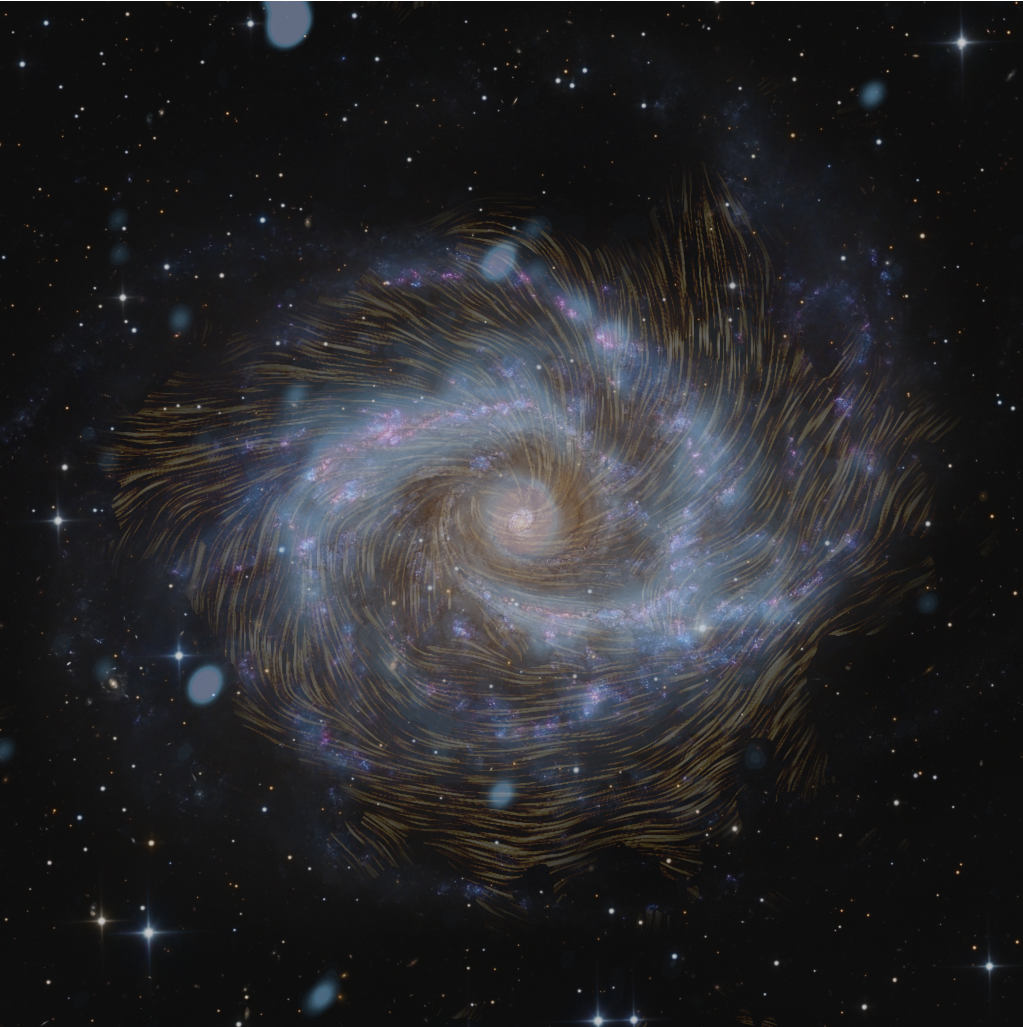}
        \caption{Magnetic field flow orientations at a resolution  of $32\arcsec\times26\arcsec$, over-plotted onto an optical image of NGC\,2997 and the total intensity radio emission (blue areas). The total radio intensity was obtained using a robust 0 weighting during the cleaning process and shows a resolution of 12.8" x 9.8". The flow was created using the licplot code (https://github.com/alexus37/licplot.git). The optical map was imaged in LRGBHa on a Planewave CDK 1000 at Observatorio El Sauce, Obstech, Chile (courtesy of Michael Selby).}
        \label{fig:optical_radio}
\end{figure*}

We did not observe any effect coming from a large-scale vertical field that would distort the sinusoidal RM variations in the azimuthal angle.
Observations of edge-on spiral galaxies confirm this view. None of the edge-on galaxies from the CHANG-ES sample observed with the VLA in the C band (5\,GHz) \citep{Irwin_2012} show any large-scale RM pattern (see Appendix of \cite{Krause_2020}). Instead, the directions of the (tilted) vertical fields reverse on scales of a few kpc, as in NGC\,4631 \citep{Mora_2019}. Recent VLA S-band observations of the edge-on galaxy NGC\,3556 by \citet{Xu_2025} allowed for the detection of polarized emission and RMs out to larger heights above the plane than with C-band data.

Vertical fields are best visible in RM maps of almost face-on galaxies, but
no large-scale vertical fields were detected so far.
Vertical fields with reversing directions on kpc scales are seen in the RM data of IC\,342 \citep{Beck_2015_IC},
NGC\,628 \citep{Mulcahy_2017}, and M\,51 \citep{Kierdorf_2020}. Such vertical fields would just add small-scale fluctuations to the Fourier transforms of RMs.

\section{Conclusions}
\label{sec:conclusions}

Taking advantage of the new capabilities of the MeerKAT radio telescope with S-band receivers, a series of observations were conducted to showcase its enhanced capabilities. This analysis focuses on the radio observations of the galaxy NGC\,2997 in full Stokes polarization. Polarization observations in the S band are ideal for studying magnetic fields in spiral galaxies. In this frequency band, Faraday depolarization effects are less severe than at lower frequencies and, due to the steep synchrotron spectrum, we are able to detect larger flux densities than at higher frequencies.

The data calibration procedure utilized the Max Planck MeerKAT Galactic Plane Survey (MMGPS) pipeline, which includes comprehensive full-Stokes calibration, automated self-calibration, and imaging capabilities tailored for L- and S-band MeerKAT data. The resulting radio maps exhibit an rms noise of $11\,\mu\rm{Jy}\,\rm{beam}^{-1}$ at a resolution of about $4\arcsec$, enabling a detailed examination of the galaxy. The total radio intensity reveals the spiral structure associated with the star-forming regions and the inner ring encircling the nucleus. Additionally, the galaxy exhibits strong, extended polarized emission indicative of a large-scale ordered magnetic field. The polarization aligns with the inter-arm regions, faithfully tracing the spiral structure of the disk.

Our RM synthesis analysis elucidates the direction of the magnetic field along the line of sight throughout the entire disk. Leveraging the sensitivity and high resolution provided by the MeerKAT's S-band capability, this study achieves an unprecedented level of detail. This demonstrates the power of S-band polarization observations with MeerKAT. 

A sector-based analysis of the RMs with azimuthal angle in the galaxy plane unveils periodic patterns in eight rings between 0.5\,kpc and 16\,kpc, suggesting the existence of azimuthal modes of the large-scale spiral magnetic field in the disk of NGC\,2997.

The variations seen for the RMs along the azimuthal angle reveal smoothly changing phase shifts between the rings. This disproves the claim of a large-scale field reversal at about 3\,kpc radius between the central region and disk by \citet{Han_1999}, leaving M\,31 and IC\,342 as the only candidates for such a reversal (see Table~1 in \citet{Beck_2019}).

Further refinement involves computing RMs modified for the effect of inclination of the disk and considering the position angle of the major axis. We suggest that this correction effectively mitigates the impact of the inclination on the component of the magnetic field parallel to the line of sight.

For the first time, a Fourier analysis was applied to RM data averaged in sectors of rings in the disk of a spiral galaxy. The RM averages for seven rings at distances between 2 kpc and 16 kpc from the center of NGC\,2997 display three distinct Fourier peaks, indicating the presence of three azimuthal magnetic field modes in the galaxy. After a geometric modification of the RM distribution, the positions of the second and third Fourier peaks shift from odd to even multiples of the position of the first peak. Dynamo modes of an even order (i.e., 0, 2, and 4) are generally aligned
with the predictions of dynamo models for a spiral galaxy with a symmetric four-arm pattern. However, the unexpected strength of the m=4 mode warrants further study.
\begin{acknowledgements}
The MeerKAT telescope is operated by the South African Radio Astronomy Observatory, which is a facility of the National Research Foundation, an agency of the Department of Science and Innovation.

This work has made use of the “MPIfR S-band receiver system” designed, constructed, and maintained by funding of the MPIfR and the Max-Planck-Society.

In particular, we acknowledge the contribution and efforts of the experts of the Electronics and Digital Signal Processing departments of the MPIfR in this project.

We acknowledge the computing support at the GLOW HPC in the Forschungszentrum J\"ulich.

M.R.R. is a Jansky Fellow at the National Radio Astronomy Observatory.

OMS's research is supported by the South African Research Chairs Initiative of the Department of Science and Technology and National Research Foundation (grant No. 81737). --
We thank Jack Livingston for his valuable comments and Luke Chamandy for sharing insights into dynamo models.

We thank the anonymous referee for all the helpful comments and suggestions, which improved the clarity and interpretation of the manuscript.

We acknowledge the contribution of our dear colleague and collaborator Karl Menten who sadly passed away far too early. 

\end{acknowledgements}

\bibliographystyle{aa}
\bibliography{bibliography}

\end{document}